\documentclass[a4paper,prd,aps,preprint,tightenlines,superscriptaddress,nofootinbib,showpacs,showkeys,floatfix]{revtex4}
\usepackage{mathrsfs}
\usepackage{amsfonts}
\usepackage{array}
\usepackage{epsfig}
\usepackage{rotating}
\usepackage{multirow}

\usepackage{amsmath}    
\usepackage{verbatim}   
\usepackage{color}      
\usepackage{colordvi}
\usepackage{bm}

\usepackage{subfigure}  
\usepackage{url}
\usepackage[colorlinks = true,
            linkcolor = blue,
            urlcolor  = blue,
            citecolor = blue,
            anchorcolor = blue]{hyperref}
\usepackage{doi}

\usepackage{pstricks,rotating, graphicx}
\usepackage{cancel}

\usepackage{tablefootnote}  

\graphicspath{{./fig/}}

\usepackage{pslatex}
\usepackage{txfonts}
\usepackage[latin1]{inputenc}
\usepackage[T1]{fontenc}

\begin{document}
\begin{titlepage}
\thispagestyle{empty}
\noindent
INR-TH-2023-012\\
\hfill
July 2023 \\
\vspace{1.0cm}

\begin{center}
  {\bf \Large
Impact of SeaQuest data on PDF fits at large $x$
  }
  \vspace{1.25cm}

 {\large
   S.~Alekhin$^{\, a}$,
   M. V. Garzelli$^{\, a}$,
   S.~Kulagin$^{\, b}$,
   and
   S.-O.~Moch$^{\, a}$
 }
 \\
 \vspace{1.25cm}
 {\it
   $^a$ 
II. Institut f\"ur Theoretische Physik, Universit\"at Hamburg \\
   Luruper Chaussee 149, D--22761 Hamburg, Germany \\
   \vspace{0.2cm}
   $^b$ Institute for Nuclear Research of the Russian Academy of Sciences \\ 117312 Moscow, Russia
 }

\vspace{1.4cm}
\large {\bf Abstract}
\vspace{-0.2cm}
\end{center}
We evaluate the impact of recent SeaQuest (FNAL-E906 experiment) data on dimuon production in
proton-deuteron and proton-proton collisions on parton distribution functions (PDFs). 
We find these data in a good agreement with the QCD predictions based on PDFs fitted 
to the Tevatron and LHC data on forward production of $W$ and  $Z$ bosons.
As a basis for this study we use the ABMP16 PDF fits and show that they turn
out to be compatible with the SeaQuest data, 
and that these data have constraining power, allowing to reduce the uncertainties 
on the isospin asymmetry of the light-sea-quark distribution at large longitudinal momentum fraction $x$. 
We discuss the nuclear corrections needed to describe the deuteron and 
show that they affect the theoretical description of the proton-deuteron Drell-Yan cross section
at the level of $\mathcal{O}(0.5 - 1)$\%.
We also comment on the compatibility of the SeaQuest results with other
state-of-the-art PDF fits and show that these data are in clear disagreement with models proposing an SU(3)-flavor symmetric quark sea.
Finally, we perform a comparison between the second Mellin moments of the light-quark PDFs and recent results from various lattice QCD computations, 
which demonstrates good compatibility, albeit limited by the uncertainties inherent in current lattice QCD simulations.

\end{titlepage}

\newpage
\setcounter{footnote}{0}
\setcounter{page}{1}

\section{Introduction}
\label{sec:intro}

The knowledge of parton distribution functions (PDFs) at large longitudinal
momentum fraction $x$ is one of the most urgent open
questions~\cite{Accardi:2016ndt, Begel:2022kwp}  concerning proton and nuclear structure to which not only theoretical but even experimental efforts are going to be dedicated in the future.
While in the long term data from the Electron Ion Collider (EIC)~\cite{AbdulKhalek:2021gbh, Abir:2023fpo} are expected to play a very important constraining role, as also emphasized in the Snowmass 2021 EIC-dedicated whitepaper~\cite{AbdulKhalek:2022hcn}, in the near future further experiments might also offer promising opportunities. 
Among those at the Large Hadron Collider (LHC), we mention here the fixed-target (FT) con\-fi\-gu\-rations exploiting one of the LHC
beams~\cite{Hadjidakis:2018ifr}, a possibility already realized by
complementing the LHCb detector with the SMOG and SMOG2 apparata~\cite{Bursche:2018orf, DiNezza:2651269}, and also conceptually
studied, although not realized, by the ALICE collaboration with the ALICE-FT
experiment~\cite{Galluccio:2671944}, as well as perspective projects still under discussion, 
like the Forward Physics Facility~\cite{Anchordoqui:2021ghd, Feng:2022inv}.  
In particular, the LHCb~+~SMOG system has already delivered the first data using proton and Pb beams impinging on gaseous nuclei like $^4$He, $^{20}$Ne and $^{40}$Ar, at different nucleon-nucleon center-of-mass energies $\sqrt{s_{NN}} \sim \mathcal{O}(50 - 100)$ GeV, corresponding to various past LHC runs. 
The LHCb~+~SMOG2 system, active during Run 3 and 4, can make use of even
lighter  
gases, like deuterium $^2$H as well as hydrogen,
with increased
statistics~\footnote{The data released so far concern open and hidden charm
  production. We expect that in the future even data on Drell-Yan production
  will become available.}. 
These experiments
allow to probe the 
longitudinal momentum fraction interval $0.1 < x < 1$ for target partons, on
which the constraints from the sets of HERA data~\cite{H1:2015ubc} which traditionally form the backbone of PDF fits, are quite loose and mostly indirect~\footnote{HERA has also delivered some sets of experimental cross-section data which could constrain PDFs at large $x$, up to $\sim 1$, see e.g. Ref.~\cite{ZEUS:2013szj}, but these data have not been used in most of the PDF fits.}.

For the time being, constraints on PDFs at large $x$ are imposed by legacy measurements 
from inclusive deep-inelastic scattering (DIS) experiments at fixed-targets (SLAC, BCDMS, NMC, etc.), 
semi-inclusive DIS experiments using $\nu$ beams and capable of measuring heavy-quark production
in DIS (CCFR, NuTeV, CHORUS, NOMAD, etc.) 
and fixed-target Drell-Yan (DY) experiments (CERN-NA51, FNAL-E605, FNAL-E866, etc.), 
complemented by measurements of cross-sections for DY (+ jets) production 
and other specific processes in the main detectors at the Tevatron and the LHC in the standard collider-mode
configuration~(for an overview, see e.g. Ref.~\cite{Accardi:2016ndt} and references therein). 

The valence quark distributions are constrained by DIS HERA data, up to
$x<0.1$, and in fixed-target experiments, up to $x \sim 1$.
A large-$x$ and relatively low-$Q$ domain is also probed at JLab~\cite{JeffersonLabHallATritium:2021usd}. 
The DY data from the Tevatron and the LHC (both inclusive cross-sections and charge
asymmetries) as well as from fixed-target experiments have also been used to
probe up ($u$) and down ($d$) quark distributions and their differences (isospin asymmetries).  
Single-top quark production data have allowed to probe the $u/d$ ratio 
at $x \sim 0.1$, where $u = u_{\rm val} + u_{\rm sea}$ and $d$~=~$d_{\rm val}$~+~$d_{\rm sea}$, notwithstanding the big systematic uncertainties still accompanying the
experimental cross-sections for this channel of top-quark production~\cite{ATLAS:2014sxe}. The (anti)strange sea quark distributions ($s, \bar{s}$) has been constrained by DY (+ jets) LHC data and older (anti)-neutrino-nuclear 
DIS data, with large uncertainties~\cite{Accardi:2016ndt,Alekhin:2017olj,Faura:2020oom},
and improving their determination remains one of the pressing issues in PDF analysis.
The $s(x)-\bar{s}(x)$ asymmetry~\cite{Catani:2004nc} can be constrained by semi-inclusive DIS data on dimuon production distinguishing neutrino and antineutrino beams (as discussed e.g. in Ref.~\cite{Bailey:2020ooq} and~\cite{Hou:2022onq}), by $W^+$~+~$\bar{c}$ and $W^-$~+~$c$ data at the LHC~\cite{Bevilacqua:2021ovq} and by future DIS experiments using separate beams of neutrinos and anti-neutrinos (e.g. at the Forward Physics Facility). 
The up and down sea quark distributions are well constrained by DY data.
Finally, the gluon PDF at large-$x$ is mostly constrained 
by measurements of heavy-quark and jet production at the LHC~\cite{Amoroso:2022eow}.

Recently, the SeaQuest collaboration (FNAL-E906 experiment) has released fixed-target data on dimuon
production on $^2$H and proton targets through DY, which allow to constrain the difference between down and up sea quarks, i.e. $\bar{d}(x)-\bar{u}(x)$, 
and the $\bar{d}(x)/\bar{u}(x)$ ratio~\cite{SeaQuest:2022vwp}. 
This experiment can be considered as a continuation of previous ones,
FNAL-E866~\cite{Towell:2001nh} (NuSea) and FNAL-E605~\cite{Moreno:1990sf}, lowering the
center-of-mass energy $\sqrt{s}$ and extending the kinematic coverage in $x$~\footnote{Another past experiment providing insights on isospin symmetry breaking through ratios of $pd$ and $pp$ DY cross-sections, following the idea of Ref.~\cite{Ellis:1990ti} and using a CERN-SPS proton beam of 450 GeV/$c$, was NA51~\cite{NA51:1994xrz}. However, they gave results in the form of only one data point around $x\sim 0.18$. Their results are compatible within experimental uncertainties with those of E866 that covers a wider $x$ range.}.
Reviews on the
flavour structure of the nucleon sea, triggering further investigations, have been provided by e.g.~Ref.~\cite{Geesaman:2018ixo, Chang:2014jba}. 
The new experimental results and the present theoretical scenario motivate the present study, where we focus on the light-quark distributions, with particular emphasis on the sea quark case. 
In Sec.~\ref{sec:abmp}, we show the impact of the SeaQuest results on the ABMP16 fits, 
considering both versions, at next-to-leading order (NLO) and at next-to-NLO
(NNLO) in perturbative QCD, published in Refs.~\cite{Alekhin:2018pai} and \cite{Alekhin:2017kpj}, respectively. 
This leads to new PDF fits, dubbed as ABMP16~+~SeaQuest NLO and NNLO,
performed ab-initio using the same statistical methodology and inputs as for
the ABMP16 fits, plus the most recent SeaQuest data. 
In Sec.~\ref{sec:other} we comment on the compatibility of other
state-of-the-art PDF fits with these data and 
in Sec.~\ref{sec:nuclear} we discuss nuclear corrections.
In Sec.~\ref{sec:lattice} we compare the second moments of the light-flavor quark PDFs with recent lattice QCD results. Our conclusions are delivered in Sec.~\ref{sec:conclu}.

\section{Constraining power of the SeaQuest data on the ABMP16 NLO and NNLO PDF fits}
\label{sec:abmp}

The study extends the ABMP16 PDF fits (NLO and NNLO), 
which have used the combined data from HERA for inclusive DIS, 
data from the fixed-target experiments NOMAD and CHORUS for neutrino-induced DIS, 
as well as data from Tevatron and the LHC for the DY process and the hadro-production of single-top and top-quark pairs.
The ABMP16 approach uses a fixed-flavor number scheme for $n_f=3, 4, 5$ 
and simultaneously determines the PDFs, the value of the strong coupling
$\alpha_s(M_z)$ and all masses of heavy quarks, 
fully preserving the correlations among these quantities.

\begin{figure}[t!]
    \centerline{
    \includegraphics[width=0.9\textwidth]{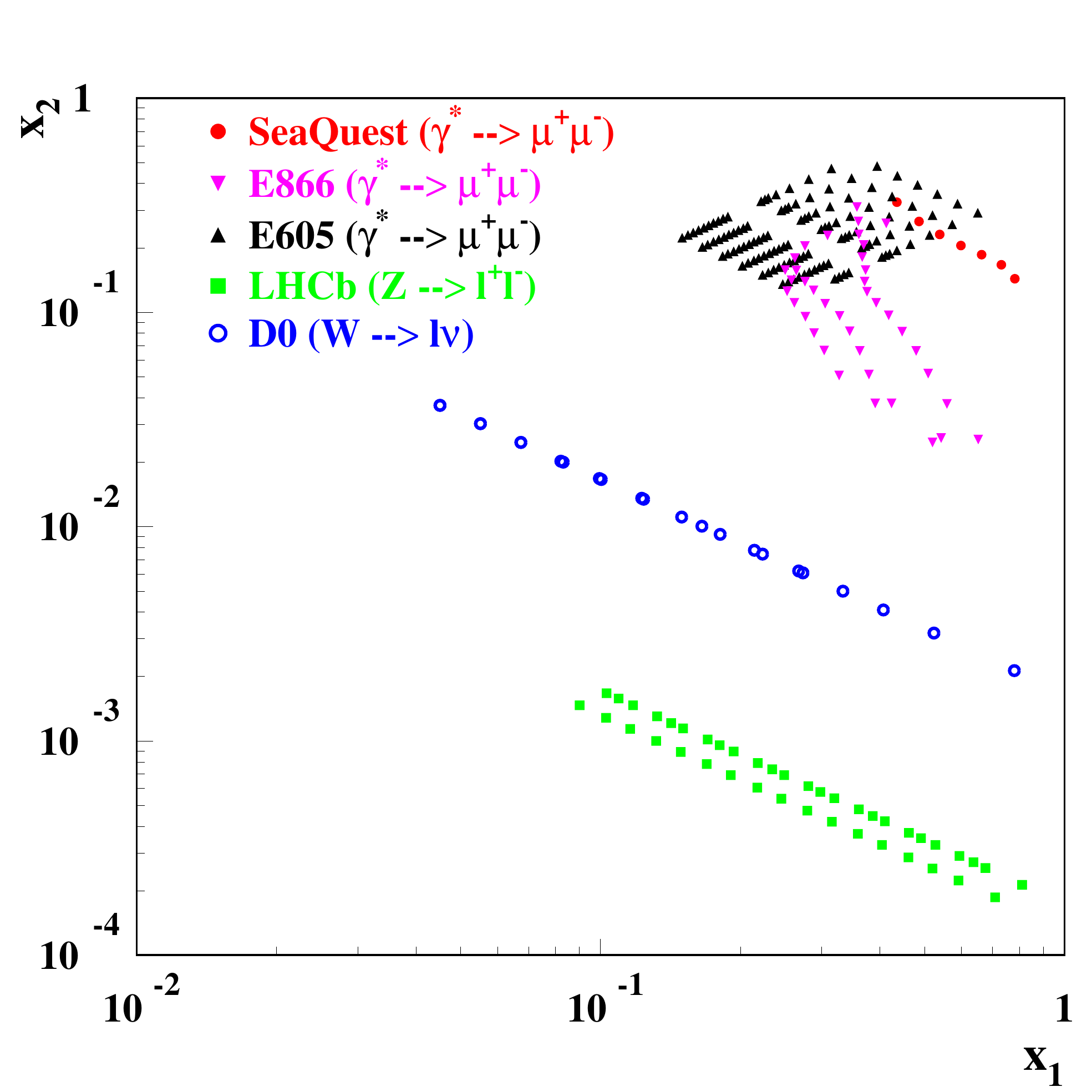}
}
\vspace*{-2mm}
  \caption{\small
  \label{fig:phase}
The ($x_1$,       $x_2$) coverage
for the SeaQuest experiment~\cite{SeaQuest:2022vwp} (full circles), with  $x_{1,2}$ 
 given by Eq.~(\ref{eq:x12}), in comparison to the coverages 
   from   DY  data of other experiments
used in the ABMP16 PDF
fits 
(down-oriented triangles:
E866 Fermilab fixed-target experiment~\cite{Towell:2001nh}; up-oriented triangles: E605~\cite{Moreno:1990sf}
with $x_{1,2}$ computed from the lepton-pair rapidity using Eq.~(\ref{eq:rap}) for both the data sets; 
squares: LHCb, the LHC experiment~\cite{Aaij:2015gna,Aaij:2015vua,Aaij:2015zlq} 
with $x_{1,2}$ computed from the $Z$-boson rapidity using Eq.~(\ref{eq:rap});
 open circles: D0, the Tevatron collider experiment~\cite{Abazov:2013rja,D0:2014kma},  
with $x_{1,2}$ estimated from the charged-lepton rapidity using Eq.~(\ref{eq:rapw})).
  }
\end{figure}

For illustrative purpose, 
we summarize in Fig.~\ref{fig:phase} the $(x_1, x_2)$ coverage of most of the DY data used in
constraining the up and down sea quark distributions at large $x$ in these
fits~\footnote{\label{foot:lhcb} We leave out the LHCb $pp \rightarrow W^\pm +  X \rightarrow  l^\pm + \overset{(-)}{\nu} + X$ production data~\cite{LHCb:2015okr, LHCb:2015mad}, whose kinematical coverage is similar to the one from LHCb $pp$~$\rightarrow$~$Z$~+~$X$~$\rightarrow$~$l^+ l^-$~+~$X$ data shown in the plot.}, together with the $(x_1, x_2)$ coverage of the recently released SeaQuest data.
The variables $x_1$ and $x_2$ represent the momentum fractions carried by the incident (anti)quarks in beam and target, respectively, which roughly characterize the region of $x$ probed by a particular experiment. 
Since $x_1$ and $x_2$ are not observables and cannot be measured, 
we detail here how we reconstruct them, assuming leading order (LO) kinematics. 
For the SeaQuest experiment $x_{1,2}$ are computed as follows:
\begin{equation}
\label{eq:x12}
x_{1,2}=\frac{P_{1,2}\cdot Q}{P_{1,2}\cdot P} \, ,
\end{equation}
where $Q$ is the four-momentum of the
virtual photon from the quark-antiquark annihilation in the non-resonant production process, $P_{1,2}$ are the four-momenta of the projectile and target hadron, respectively, and $P=P_1+P_2$. 
Considering $\gamma^* \rightarrow \mu^+ \mu^-$ decays, 
the average values for $x_{1,2}$ in the bins of the muon-pair average Feynman variable
$\langle x_{F} \rangle$ are reported in Ref.~\cite{SeaQuest:2022vwp}. These values are plotted in Fig.~\ref{fig:phase} in comparison with the kinematics of other DY data included in the ABMP16 fits. In particular, for the E605 Fermilab fixed-target data~\cite{Moreno:1990sf}, given
in the form of a double differential 
distribution in $\sqrt{\tau}=M/\sqrt{s}$ and $y$, 
where $M$ and $y$ are the invariant mass and rapidity of the $\mu^+\mu^-$-pair, respectively, and $\sqrt{s}$ is the collision center-of-mass energy, the values of $x_{1,2}$ are computed according to the relation
\begin{equation}
\label{eq:rap}
x_{1,2}=\sqrt{\tau}e^{\pm y} \, .
\end{equation}
For the E866 experiment~\cite{Towell:2001nh} the same relation is employed. However, since the muon-pair ra\-pi\-di\-ty is not 
tabulated in Ref.~\cite{Towell:2001nh}, 
it is computed from the muon-pair $x_{F}$ and transverse momentum $p_T$ 
using basic definition as follows:
\begin{equation}
\label{eq:rapb}
y=\frac{1}{2}\ln\left(\frac{E+p_L}{E-p_L}\right) \, ,
\end{equation}
where $p_L = x_F \, p_{L,\, \rm{max}}$ and $E = \sqrt{p_L^2 + p_T^2 + M^2}$ are the muon-pair longitudinal momentum and energy, respectively, in the center-of-mass frame of the colliding hadrons, with $p_{L,max}$ the maximum longitudinal momentum of the muon-pair, depending on $\sqrt{s}$ according to the formula $p_{L,max} = \sqrt{s}\,(1 - M^2/s)/2$.
  
The approach of Eq.~(\ref{eq:rap}) is also used for the LHCb data on
$Z$-boson production~\cite{Aaij:2015gna,Aaij:2015vua,Aaij:2015zlq} released in the form of lepton-pair pseudorapidity distributions~\footnote{For the lepton energies of LHCb the numerical difference between pseudorapidity and rapidity is negligible.}.
The data on $W$-boson production evidently probe the same kinematics.
However, the use of Eq.~(\ref{eq:rap}) is impossible in this case due to the neutrino escaping detection. Therefore, for $W$-boson
production 
in the D0 experiment~\cite{Abazov:2013rja,D0:2014kma},
we use the following approximate estimate:
\begin{equation}
\label{eq:rapw}
x_{1,2} = \frac{M_W}{\sqrt{s}}e^{\pm y_l} \, ,
\end{equation}
where $M_W$ is the $W$-boson pole mass and $y_l$ is the lepton rapidity.

Both DY data at the Tevatron and the LHC and DY data in fixed-target experiments
play a role in constraining the sea quark PDFs at large $x$. 
They allow to reach similarly large-$x$ values, although in the case of 
fixed-target data, relatively large-$x$ partons from both the projectile and
the target participate in the same hard interaction, whereas in the case of the LHC,
a large-$x$ parton is typically probed simultaneously with a low-$x$ one, 
as exemplified in the ($x_1$, $x_2$) correlation in Fig.~\ref{fig:phase}. 
The correlation is quite evident for the LHC data and is related to the exchange of heavy bosons in the DY process.  
In the case of LHC, the largest $x_1$ values are probed by the LHCb detector with data at large positive rapidity, which covers the 
interval $2 < y < 4.5$. 
On the other hand, the fixed-target experiments E866 and E605, which have much
lower center-of-mass energies than the LHC, probe larger $x_2$ values
and present a less evident ($x_1$, $x_2$) correlation, related to the exchange
of a $\gamma^*$ with a broad range of mass values in the DY process. 
In the case of SeaQuest, the ($x_1$, $x_2$) correlation is again evident,
considering that the invariant mass of the observed $\gamma^*$ decay products
is fixed to approximately $M \sim$ 5~GeV. SeaQuest covers $x_2$ values higher
than the LHC due to the use of a beam with much lower center-of-mass energy ($\sqrt{s} = 15.1$~GeV). 
The $x_2$ region covered by SeaQuest extends up to $\le$ 0.45. 
The E605 experiment has a coverage extending even up to slightly higher $x_2$ values.  
However, the E605 experiment used a Copper target, thus requiring an evaluation of nuclear corrections (see the end of Section~\ref{sec:nuclear}). Having only one target material, they could not provide data on cross-section ratios, unlike  SeaQuest that has both a deuteron and a proton target. Also, given that
Copper is a heavy nucleus close to isoscalarity, the E605 data are much less sensitive to the isospin asymmetry effects, that we investigate in this work. 
We explicitly verified the very small impact of E605 data, by removing them from our fits where they are included as default.

The green dots along two parallel lines in Fig.~\ref{fig:phase} refer to the
cases of $Z$-boson production at the LHC at $\sqrt{s}$~=~7 and 8 TeV, given 
that data at these center-of-mass energies were included in the ABMP16 PDF fits. 

\begin{figure}[h!]
    \centerline{
    \includegraphics[width=0.9\textwidth]{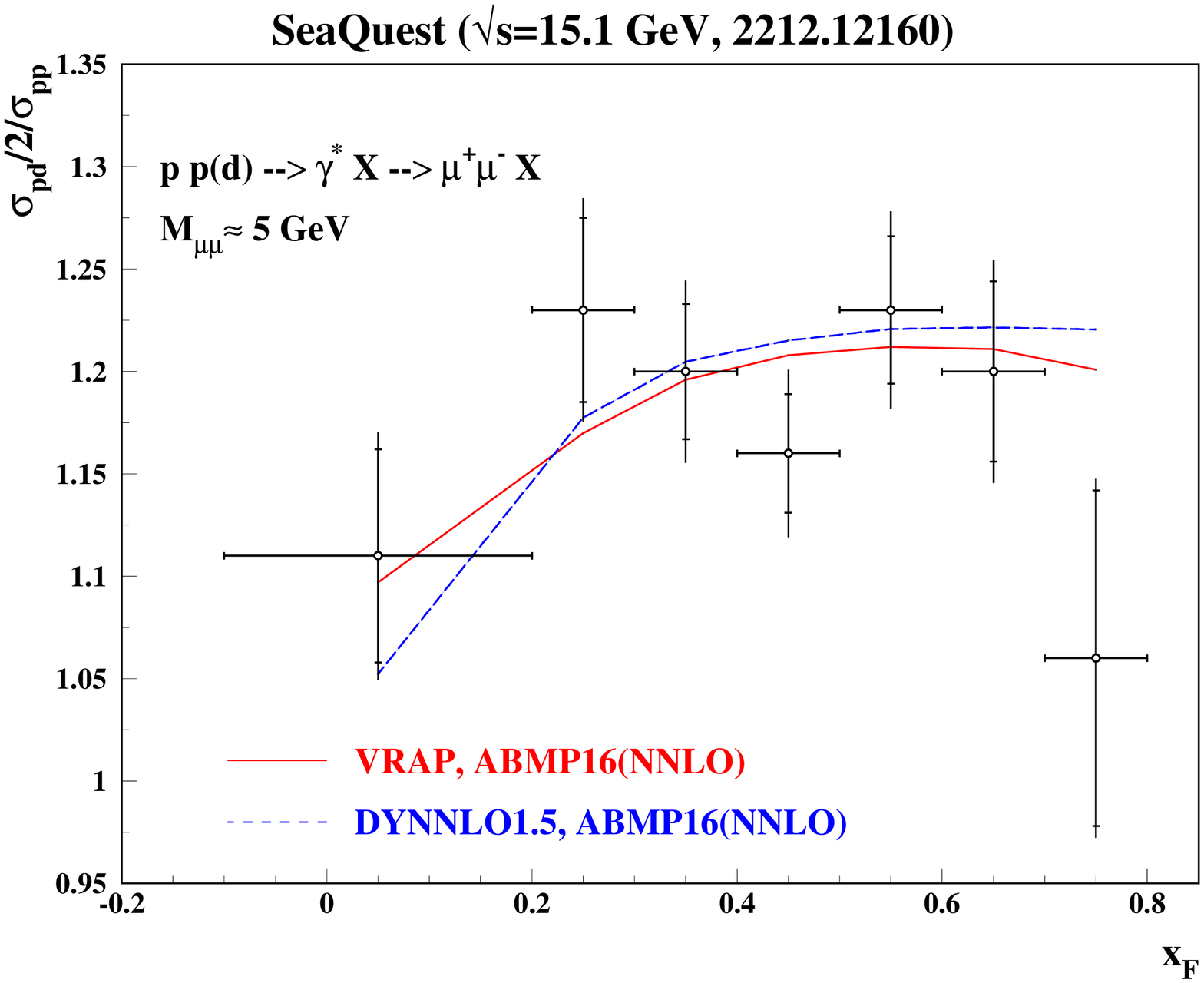}
}
\vspace*{-2mm}
  \caption{\small
  \label{fig:dy}
The SeaQuest data~\cite{SeaQuest:2022vwp} on the ratio of $pd$ and $pp$ DY distributions over Feynman variable $x_F$
with respect to the NNLO predictions obtained using the code VRAP~\cite{Anastasiou:2003ds} (solid line) and DYNNLO~\cite{Catani:2009sm} (dashes). 
}  
\end{figure}

In order to compute predictions for the DY cross-sections, we use the 
FEWZ2.1 code~\cite{Gavin:2012sy} for the collider cases and the 
VRAP code~\cite{Anastasiou:2003ds} for the fixed-target cases. 
In particular, the present analysis of SeaQuest data is based on the $x_F$-distribution, that was directly measured in the ex\-pe\-ri\-ment and could also be computed 
to NNLO QCD accuracy using a Monte-Carlo code, like e.g. FEWZ or
DYNNLO~~\cite{Catani:2009sm}~\footnote{Note that NNLO predictions for the 
  DY process obtained using non-local subtraction methods, 
  may differ by power corrections, whose size varies depending on the experimental
  cuts on final-state leptons~\cite{Alekhin:2021xcu}. 
  These corrections are, however, negligible in the context of this work.}. 
However, in the fit we employ the VRAP code, which is based on 2-dimensional
integration that allows to greatly improve the code performance. 
To compute VRAP predictions for the SeaQuest data on the $x_F$-distribution we
perform a mapping of $x_F$ to the rapidity using the basic relation
Eq.~(\ref{eq:rap}) and taking 
$P_L = \langle x_F \rangle \sqrt{s}/2(1- \langle M \rangle^2/s)$, 
$E = \sqrt{\langle M \rangle^2 + P_L^2 + \langle P_T \rangle^2}$, 
where $s$ is the center-of-mass energy squared and 
$\langle x_F \rangle$, $\langle M \rangle$ and $\langle P_T \rangle$ are the
averages of muon-pair Feynman variable $x_F$, invariant mass and transverse
momentum, respectively, over the bins in $x_F$. 
These averaged quantities are all given in Ref.~\cite{SeaQuest:2022vwp} for
each bin in $x_F$.  
To validate such an approach, we compare its predictions with those obtained
with the methodology used in Ref.~\cite{SeaQuest:2022vwp}, where the
DYNNLO~\cite{Catani:2009sm} code is employed, instead of VRAP, and the exact
information con\-cer\-ning the transverse momentum and the invariant mass of the
$\mu^+\mu^-$-pair on an event-by-event basis is considered to build the $x_F$ distributions, 
instead of the average value of these quantities per $x_F$ bin. 
We find that the difference is mostly well below the data uncertainties, cf. Fig.~\ref{fig:dy}, where we compare predictions obtained with VRAP using the approximations
outlined above with the DYNNLO predictions based on the exact values for $P_T$ and
$M$ as input, as in Ref.~\cite{SeaQuest:2022vwp}, and applying their same cut $M >$~4.5~GeV. 
The latter suppresses the $\mu^+\mu^-$-background contribution from $J/\psi$ and
$\psi^\prime$ production and decay. 
We build the $x_F$ distributions using Eq.~(4) of Ref.~\cite{SeaQuest:2022vwp}. 
From Fig.~\ref{fig:dy} it is evident that only in the smallest $x_F$ bin the
difference between VRAP and DYNNLO is comparable to the data uncertainty, 
an observation that might be related to the width of the bin, which is much
larger for this bin, than for the other ones. 
Obviously, such a difference cannot have relevant impact on fit results. 
Therefore, considering that the approximated procedure with the use of VRAP allows for
NNLO simulations much faster than the exact procedure using DYNNLO, and given
that the results turn out to be very well compatible, we use VRAP for the analyses and
all other plots presented in the rest of this work.

We observe that the corrections related to spectrometer acceptance as a
function of $x_1$ and $x_2$ reported in Ref.~\cite{SeaQuest:2021zxb} do not
impact distributions depending on measured quantities, like e.g., $x_F$, that
we consider in this work. On the other hand, their inclusion is relevant for
the extraction of the $\bar{d}-\bar{u}$ asymmetry and the 
$\bar{d}/\bar{u}$ ratio 
from the SeaQuest data, as performed by the SeaQuest collaboration in approximated form as described in their papers~\cite{SeaQuest:2021zxb, SeaQuest:2022vwp}. 

The constraints from the SeaQuest experiment turn out to be compatible with
those already imposed by collider data, as shown by the fact that the
$\chi^2/\mathrm{NDP}$ of the analyses including also the SeaQuest data does not
change significantly with respect to the $\chi^2/\mathrm{NDP}$ of the original ABMP16 analyses.
Here $\mathrm{NDP}$ indicates the number of data points and the differences are well
within the $\chi^2$ statistical uncertainties, as shown in Tab.~\ref{tab:datatot}.
The $\chi^2/\mathrm{NDP}$ for the NNLO analyses turns out to be $1.18$, slightly 
closer to 1 than the $\chi^2/\mathrm{NDP}$ of the NLO analyses, which is equal to $1.20$. 
We also observe that incorporating SeaQuest data in the fits has a negligible
impact on the values of $\alpha_s(M_Z)$ and heavy-quark masses, extracted
simultaneously to PDFs in all the fits considered in Tab.~\ref{tab:datatot}.  

Separate $\chi^2$ values for various data sets included in our NNLO QCD analyses   are reported in Tab.~\ref{tab:datahq}.
We have considered four variants: 
(I) the ABMP16 analysis, (II) the ABMP16 + SeaQuest analysis, as well as (III)
an analysis, where we consider all data of (II), except the D0 DY
data and (IV) an analysis, where we consider all data of (II), except the LHCb DY data. We include variants (III) and (IV) due to the fact that in the past we have observed some tension between the D0 and LHCb DY data. 
By comparing (I) and (II), we find that, for each considered data set, the
addition of SeaQuest data does not introduce significant modifications of the
$\chi^2$. 
Thus SeaQuest data are well compatible with both the LHCb and the D0 DY data. 
On the other hand, by comparing (II) and (III), we find that the elimination
of the D0 DY datasets from the fit allows to improve the $\chi^2$ of the
analysis of the 7 TeV LHCb DY dataset by several units, beyond the statistical
$\chi^2$ uncertainty. Vice versa, the elimination of the LHCb DY datasets allows to improve the description of D0 data, as can be understood by comparing (II) and (IV).

The $\chi^2$ values were computed accounting for statistical and systematic
uncertainties of the SeaQuest data, assuming that systematic uncertainties are
fully correlated bin-by-bin. Detailed information concerning correlations
among the uncertainties characterizing the SeaQuest data are, however, not
available. Therefore, we also consider a variant of the fit, where the
systematic uncertainties are considered as fully uncorrelated. 
We have found that the $\chi^2$ values related to the analysis of the SeaQuest
data in both analyses are compatible within statistical fluctuations 
($\chi^2_{{corr.}}$~=~7.3 vs. $\chi^2_{{uncorr.}}$~=~5.9,
for $\mathrm{NDP}$ = 7). 
This implies that more details on the precise degree of bin-by-bin
correlations of the systematic uncertainties in the SeaQuest data, 
when available, will not modify the main conclusions of our study.  

\begin{table}[t!]
\begin{center}                   
  \begin{tabular}{|c|c|c|c|}
\hline
fit & $\mathrm{NDP}$ & \multicolumn{2}{c|}{$\chi^2$}
\\
\cline{3-4}
 & & 
NLO & NNLO \\
\hline
ABMP16 & 2861 & 3428.9  &  3377.6
\\
\hline
present analysis (ABMP16 + SeaQuest) & 2868 & 3438.4  &  3384.7
\\
\hline
 \end{tabular}
\caption{\small 
\label{tab:datatot}
{The total values of $\chi^2$ obtained for the NLO and NNLO ABMP16 fits in comparison with the ones of the present analyses, including all data already considered in the ABMP16 fits plus SeaQuest data. See text for more detail.
}}
\end{center}
\end{table}

\begin{table}[!]
\begin{center}                   
  \begin{tabular}{|l|c|c|c|c|c|c|c|c|}
    \hline
Experiment & Process & $\sqrt{s}$ (TeV) & Ref.      & $\mathrm{NDP}$& \multicolumn{4}{c|}{$\chi^2$} \\
\cline{6-9}
 &  &  &  &  & 
I & II & III & IV \\ 
\hline
SeaQuest &$pp \rightarrow \gamma^* X \rightarrow \mu ^+ \mu^- X$ &0.0151 & \cite{SeaQuest:2022vwp} & 7 & --& 7.3& 8.1 & 7.6 \\
   & $pd \rightarrow \gamma^* X \rightarrow  \mu ^+ \mu^- X$ &  &  &  &  &  & &
\\
\hline
D0 &$\bar{p}p  \rightarrow W^{\pm} X\rightarrow \mu ^{\pm} \overset{(-)}{\nu}  X$ &1.96& \cite{D0:2013xqc} & 10 & 17.6& 17.6&--&14.5
\\
\cline{2-9}
 &$\bar{p}p  \rightarrow W^{\pm} X\rightarrow e ^{\pm} \overset{(-)}{\nu}  X$ &1.96& \cite{D0:2014kma} & 13 & 19.0& 19.0&--&15.9
\\
\hline
LHCb &$pp  \rightarrow W^{\pm} X\rightarrow \mu ^{\pm} \overset{(-)}{\nu}  X$ &7& \cite{LHCb:2015okr} & 31 & 45.1& 43.9&35.0 &--
\\
   &$pp  \rightarrow Z X\rightarrow \mu ^{+} \mu ^{-} X$ &  &  &  &&&&
\\
\cline{2-9}
 &$pp  \rightarrow W^{\pm} X\rightarrow \mu ^{\pm} \overset{(-)}{\nu}  X$ &8& \cite{LHCb:2015mad} & 32 &40.0 &39.6 &38.2&--
\\
   &$pp  \rightarrow Z X\rightarrow \mu ^{+} \mu ^{-} X$ &  &  &  &&&&
\\
\cline{2-9}
  &$pp  \rightarrow Z X\rightarrow e ^{+} e ^{-} X$ &8& \cite{LHCb:2015kwa} & 17 &21.7 &21.9 &21.9&--
\\
\hline
 \end{tabular}
\caption{\small 
\label{tab:datahq}
{The values of $\chi^2$ obtained for the data sets probing the large-$x$ PDFs, which are included in various analyses
(column I: NNLO ABMP16 PDF fit~\cite{Alekhin:2017kpj}, column II: present analysis, column III: a variant of present analysis with D0 DY data excluded, column IV:a variant of present analysis with LHCb DY data excluded).
}}
\end{center}
\end{table}

\begin{figure}[h!]
    \centerline{
    \includegraphics[width=0.45\textwidth]{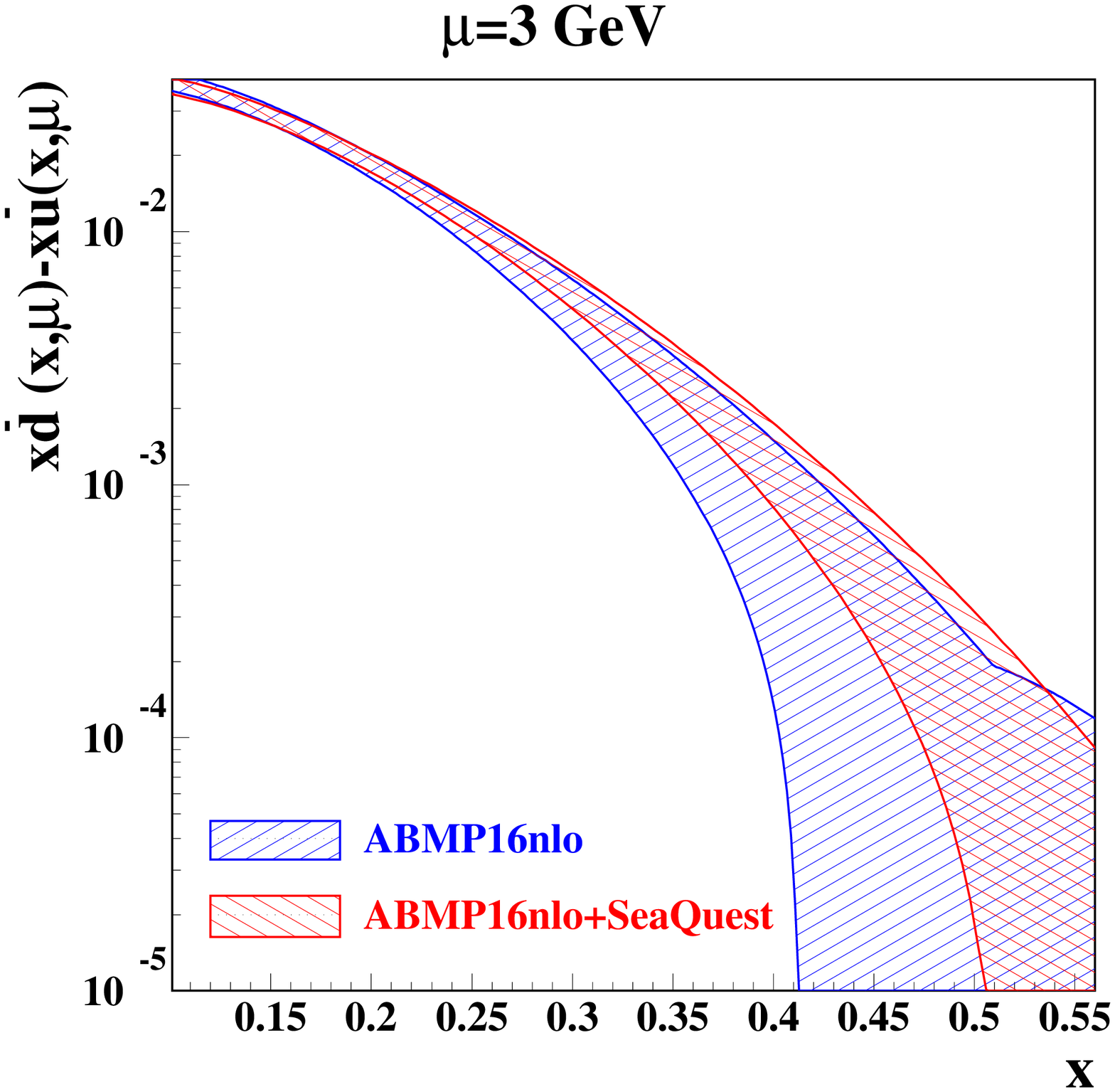}
    \includegraphics[width=0.45\textwidth]{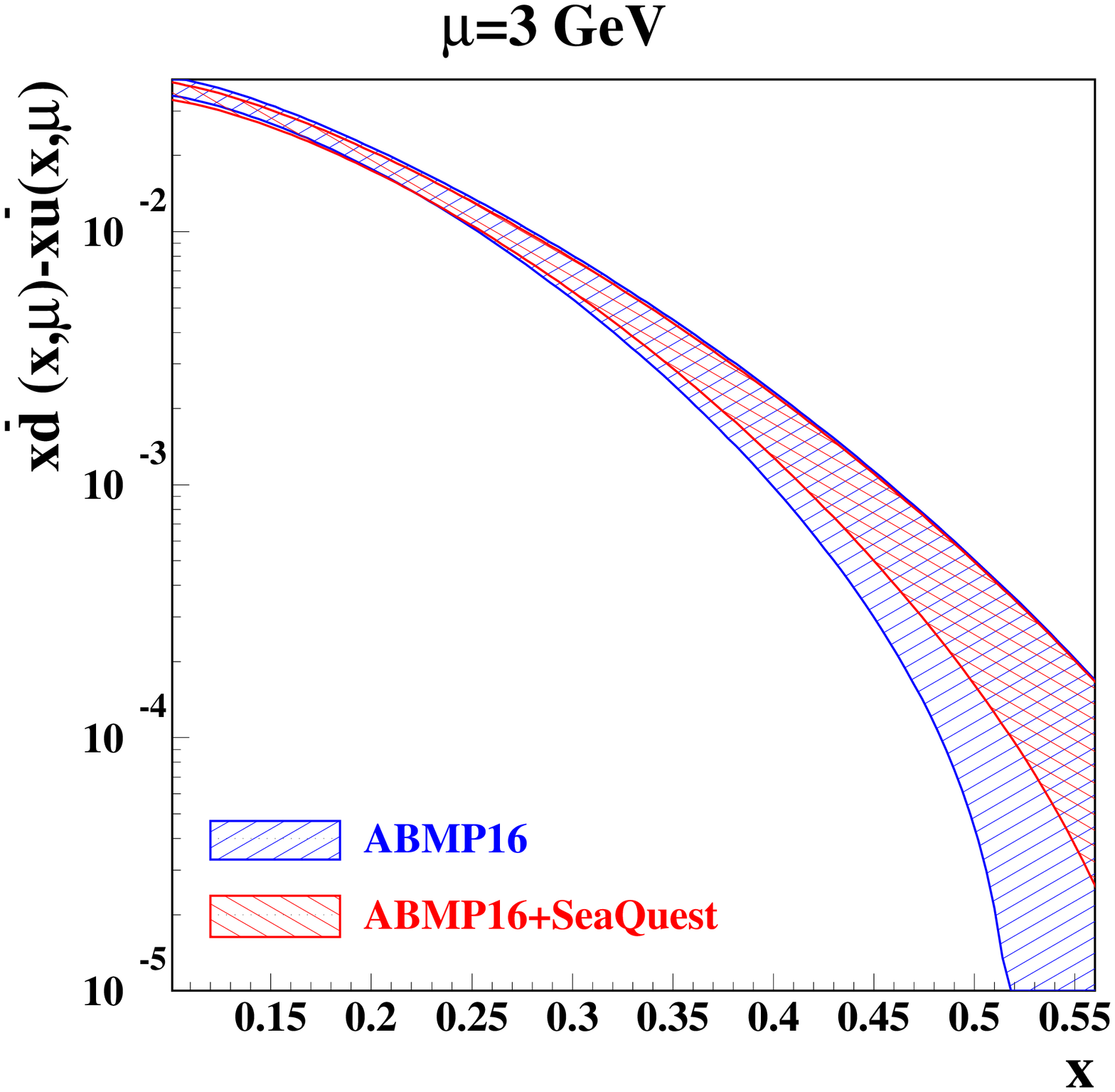}
}
\vspace*{-2mm}
  \caption{\small
  \label{fig:udm}
The $1\sigma$ band for the $n_f=3$-flavour isospin asymmetry of the sea distribution $x(\bar{d}-\bar{u}) (x)$ at the scale $\mu=$3 GeV obtained in the present analysis (left-tilted hash) compared to the one of the ABMP16 fit (right-tilted hash). The left panel shows results of the NLO analysis, whereas the right panel refers to the NNLO one. 
}
\end{figure}

\begin{figure}[h!]
    \centerline{
    \includegraphics[width=0.45\textwidth]{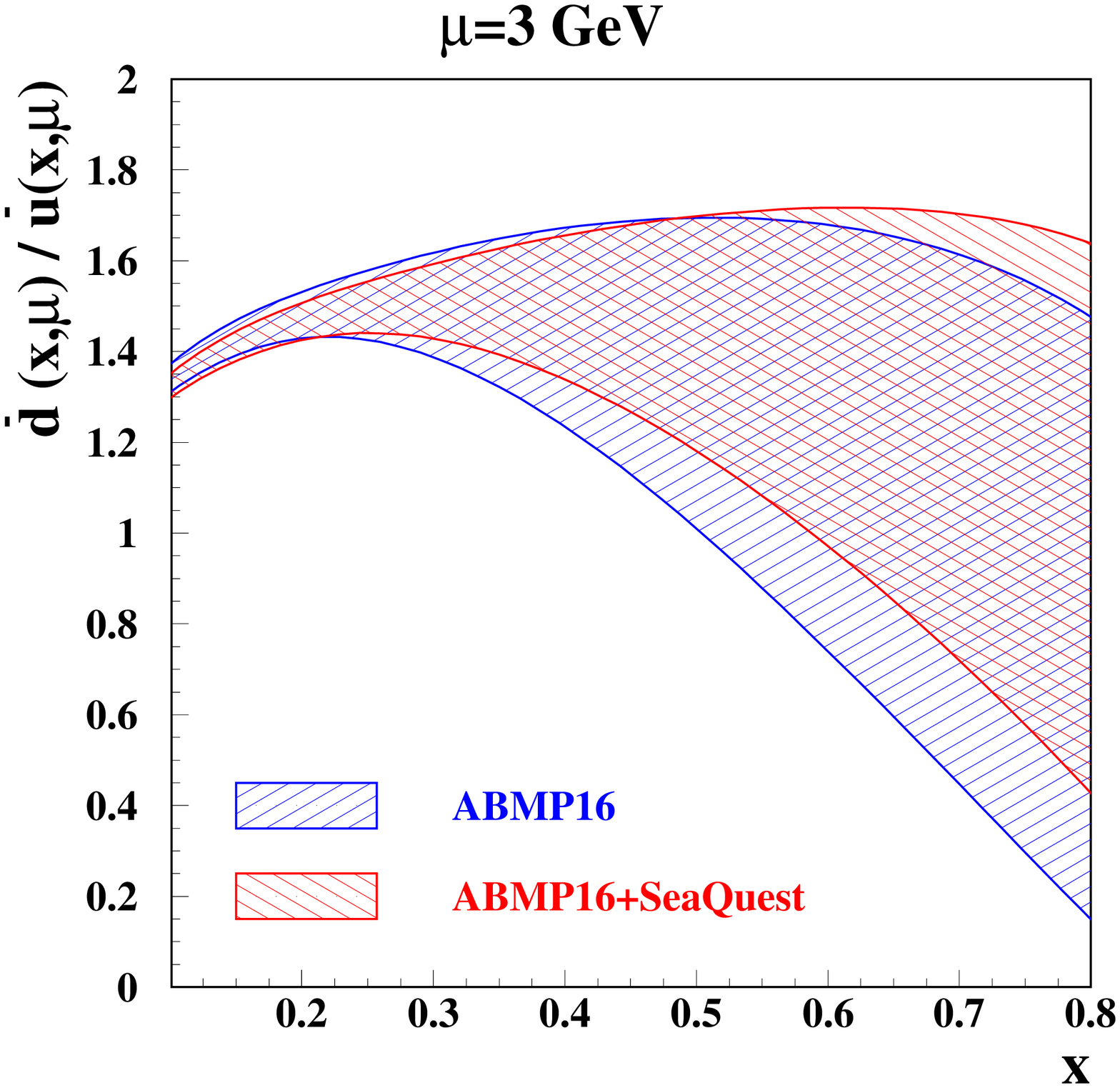}
    \includegraphics[width=0.45\textwidth]{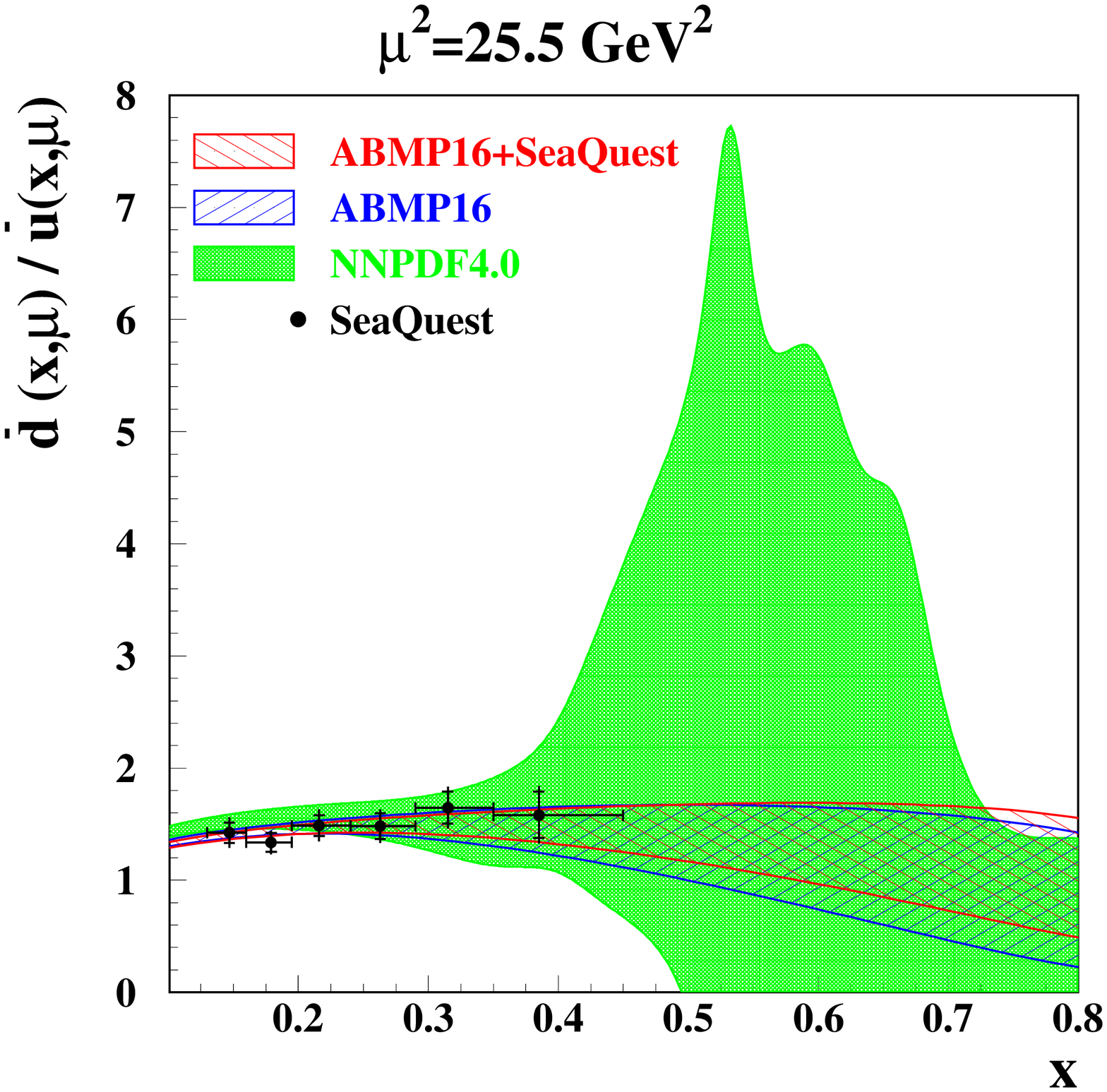}
}
\vspace*{-2mm}
  \caption{\small
  \label{fig:udmratio}
Left panel: the $1\sigma$ band for the ratio of the $n_f=3$-flavour sea distributions $\bar{d}/\bar{u}$ as a function of $x$ at the scale $\mu = $ 3 GeV obtained in the present NNLO analysis (left-tilted hash) compared to the one of the ABMP16 fit (right-tilted hash). Right panel: same as in the left panel, but at the scale $\mu^2=$ 25.5 GeV$^2$ at which the SeaQuest collaboration extracted the $\bar{d}/\bar{u}$ ratio, that is also plotted. Also shown are the 
$1\sigma$ predictions with the NNPDF4.0 NNLO PDF fit, which has incorporated SeaQuest data. 
}
\end{figure}

Fig.~\ref{fig:udm} shows the constraining power of the
SeaQuest data on the $\bar{d}(x)~-~\bar{u}(x)$ difference, in\-cre\-asing towards
large $x$ values.
At NLO, the uncertainty band of the analysis with SeaQuest data has a large
overlap, but is not completely included within the band of the default ABMP16
analysis (not including these data). 
Additionally, for $0.1 < x < 0.2$, the band of the analysis with SeaQuest data 
turns out to be 
of the same size of
 the band of the one without these data. 
On the other hand, at NNLO, the uncertainty bands are in general smaller than 
at NLO and the one of the analysis with SeaQuest data is always included and 
smaller than the band of the analysis without SeaQuest data. 
These findings confirm that theory predictions at NNLO accuracy are in general
more robust and consistent among each other than NLO ones, 
i.e., the theory description at NLO is still incomplete and hardly provides a simultaneous excellent description of all DY data, like is instead happening at NNLO. 
This is also reflected in the comparison of the $\chi^2/\mathrm{NDP}$ values presented in Tab.~\ref{tab:datatot}. 
In any case, in Fig.~\ref{fig:udm}, the constraining power of SeaQuest data is
certainly evident for $x > $ 0.3 for both the NLO and NNLO analyses. 
However, for large $x$ values the difference between the distributions of $\bar{d}(x)$ and $\bar{u}(x)$ diminishes.

Analogous observations can be made when examining Fig.~\ref{fig:udmratio},
whose left panel illustrates the variation of the ratio
$\bar{d}(x)/\bar{u}(x)$ with respect to $x$ for $\mu$~=~3~GeV. 
The ratio is larger than unity within a large $x$ interval, up to at least $x <$ 0.5 - 0.6. 
At these $x$ values, both the $x\bar{u}$ and $x\bar{d}$ sea distributions are
tiny, of the order of $10^{-5}$.  
The analysis incorporating SeaQuest data exhibits a high level of
compatibility with the analysis that excludes them, and displays a smaller
uncertainty band, especially for $x > 0.3$.
This confirms the constraining role of the SeaQuest data. 
As shown in the right panel of Fig.~\ref{fig:udmratio}, the results are also
very well compatible with the $\bar{d}(x)/\bar{u}(x)$ ratio extracted
by the SeaQuest collaboration at the scale $Q^2$~=~25.5~GeV$^2$, 
which is characteristic of the kinematics of the experiment, using as a starting point
the experimentally measured cross-section ratio $\sigma_{pd}/(2\sigma_{pp})$
and Eqs.~(8), (10) and (11) of Ref.~\cite{SeaQuest:2022vwp}. 
Although this extraction depends in principle on the PDFs used 
(the quoted SeaQuest values are those reported in Tab.~8 of Ref.~\cite{SeaQuest:2022vwp}, obtained using
cross-sections computed with the CT18 PDF fit as input of their Eq.~(11)),
this dependence is quite weak, i.e., it comes from subleading terms in Eq.~(11) of Ref.~\cite{SeaQuest:2022vwp}, 
generating minor corrections to the leading result corresponding to the case $x_1 \sim x_2$.
Therefore the extracted $\bar{d}/\bar{u}$ ratio can be considered as a robust
quantity, as also already observed in Ref.~\cite{SeaQuest:2022vwp}~\footnote{We refrain from
  comparisons with the $(\bar{d} - \bar{u})(x)$ values also reported in Tab.~8
  of Ref.~\cite{SeaQuest:2022vwp}, because these values are indeed more
  sensitive to the PDF used in their extraction, depending on additional
  assumptions on the $\bar{u}(x)$ distribution.}. 
We also note that the Seaquest data cover target $x$ values up to 0.45. 
The uncertainty band of the ABMP16~+~SeaQuest PDFs remains small at even larger
$x$ values, which is a consequence of assumptions about the parameterization
of these PDFs and their extrapolation to large $x$, 
performed under assumption of smoothness of the distributions. 
The same is true for the ABMP16 PDF fit. 
Only future experimental data in the large $x$ region will be able to check the
correctness of this extrapolated result shown here.
Regardless, it is important to emphasize that the ABMP16~+~SeaQuest fits rely on
the identical PDF parameterization employed in the original ABMP16
fits. Remarkably, this parameterization already yielded a satisfactory fit to
the new data, without necessitating any post-adjustments through the
introduction of additional parameters. 
During the original ABMP16 fit, we employed a strategy that involved
investigating the impact of various functional forms while minimizing the
number of parameters used. Our aim was to avoid introducing any additional
parameters that did not contribute significantly to an improved description of
the data.

We also point out that the effect of SeaQuest data, when comparing the ABMP16
PDFs to the ABMP16~+~SeaQuest PDFs, is not dramatic, because the ABMP16 fits
already included the E866 data, capable of constraining the $\bar{d}/\bar{u}$
ratio up to slightly lower $x$ values than SeaQuest. The main addition of
SeaQuest has been to have provided reliable measurements in the interval $x
\sim$ 0.24 - 0.45, which have helped to further constrain PDFs with respect to the past. 

\section{Compatibility of SeaQuest data with other PDF fits}
\label{sec:other}

\begin{figure}[h!]
    \centerline{
    \includegraphics[width=0.9\textwidth]{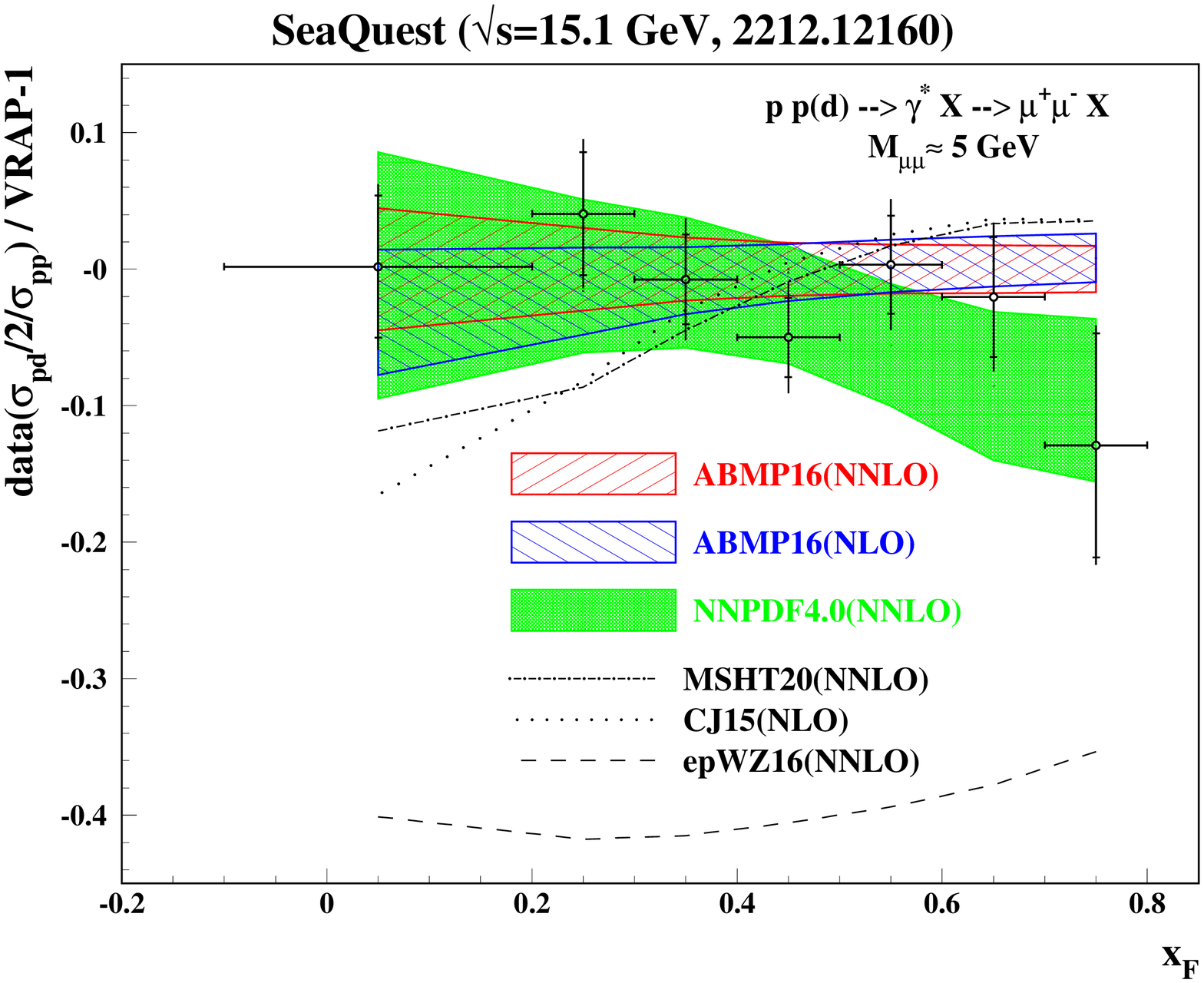}
}
\vspace*{-2mm}
  \caption{\small
  \label{fig:seaquest}
The pulls for SeaQuest data~\cite{SeaQuest:2022vwp} on the ratio of $pd$ and $pp$ DY distributions over $x_F$
with respect to predictions obtained using the code VRAP~\cite{Anastasiou:2003ds} in combination with the NNLO ABMP16 PDFs.
The $1\sigma$ band for prediction (right-tilted hash) is compared to the NLO ABMP16~\cite{Alekhin:2018pai} (left-tilted hash) and 
NNLO NNPDF4.0~\cite{NNPDF:2021njg} (shaded area) ones.  
The central values of  predictions with other PDFs are shown for comparison
(dots: NLO CJ15~\cite{Accardi:2016qay}, long dashes: NNLO epWZ16~\cite{ATLAS:2016nqi}, suggesting the SU(3)-symmetric quark sea, dashed
dots: NNLO MSHT20~\cite{Bailey:2020ooq}).   
}
\end{figure}

The compatibility of the SeaQuest data with a number of modern PDF fits
is shown in Fig.~\ref{fig:seaquest}. 
The SeaQuest data align well
with the predictions based on the NNPDF4.0 fit~\cite{NNPDF:2021njg}, which is
not surprising since the NNPDF collaboration incorporated these data into
their fitting process.
Nevertheless, the uncertainty range associated with this particular fit
remains larger compared to our own uncertainty range, in contrast to the uncertainties accompanying the data.
We argue that this behaviour can be ascribed to inefficiencies in the statistical estimators used in their analysis. 

This issue seems to be confirmed also by the predictions on the $\bar{d}/\bar{u}$ 
ratio shown in Fig.~\ref{fig:udmratio} (right), 
where the constraining power of the SeaQuest data seems to be only partially
reflected in the NNPDF4.0 uncertainties. 
This is particularly visible in the region $x \sim$ 0.3 - 0.45, 
where the NNPDF4.0 uncertainties become large, although this region is still covered by SeaQuest data. 
The inclusion of the FNAL-E605 data in the NNPDF4.0 fit should have 
imposed additional constraints on the $\bar{d}/\bar{u}$ ratio at these specific values of $x$. 
Consequently, one would expect the size of their 1$\sigma$ band to be smaller compared to the result from their fit. 
The use of a large fixed number of parameters in the parameterization of these
PDFs might be responsible for a relatively large uncertainty in the $x$ region where SeaQuest data are present.
The shape of the spike region around $x \sim 0.5$ in Fig.~\ref{fig:udmratio}
(right) seems to be driven by the step functions used in the parameterization
of these PDFs.

The large uncertainties of NNPDF4.0 at larger $x$ values in Fig.~\ref{fig:udmratio} (right), 
on the other hand, can be attributed to the lack of data. 
The smaller uncertainty of the
ABMP16 fits (in comparison to NNPDF4.0) in the very large $x$ region, not covered by the SeaQuest data, is neither related to the use of looser $W$ cuts ($W >$ 2 GeV) on the invariant mass of the hadronic system
\begin{eqnarray}
  \label{eq:wdef}
  W^2 = M_P^2 + Q^2 (1-x)/x
  \, ,
\end{eqnarray}
in DIS data~\footnote{In fact electron DIS data are sensitive to the total 
$q  = q_{\rm val} + q_{\rm sea}$ distributions, which can be separated into the valence and
  sea components only by adding DY data and/or neutrino DIS data to PDF fits.} ($M_P$ is the proton mass),  
nor to the inclusion of higher-twist corrections in the fit. 
Moreover, it is important to note that these uncertainties cannot be
considered highly informative. This is due to the fact that the uncertainty
arises solely from the extrapolation beyond the region where data are
available, relying on assumptions of smoothness, 
as already mentioned in the previous section.
It is however true that the sum rules play a role in constraining the shape of PDFs there.   
We have checked that a shape with even more spikes and larger uncertainties
for the $\bar{d}/\bar{u}$ ratio occurs in the case
of the predecessor of NNPDF4.0 fit, i.e. the NNPDF3.1 PDF fit~\cite{NNPDF:2017mvq} (not shown in the plot), not including SeaQuest data.  

These concerns about the NNPDF4.0 PDFs at large $x$ are also manifest
in
unusual
predictions for the
forward-backward asymmetry $A_{FB}^*$ in the invariant mass of the dilepton
final state at the LHC, quite different from those of many others PDF fits particularly for large invariant masses~\cite{Fiaschi:2022wgl,Ball:2022qtp}. 
The measurement of this quantity and its comparison with theory predictions might be important for
improved fits of large-$x$ quark PDFs within the Standard Model (SM) and/or for
discovering new physics associated to new gauge sectors beyond the SM, such as a heavy
neutral $Z^\prime$-boson, see, e.g. Refs.~\cite{Accomando:2017scx,Fiaschi:2021sin}.

The SeaQuest data are also compatible with the CT18 fit~\cite{Hou:2019efy}
(not shown in Fig.~\ref{fig:seaquest}), although the uncertainty of the latter
looks particularly large, even due to the tolerance criterion used in this PDF
fit ($\Delta \chi^2 = 100$ at 90\% C.L. roughly corresponding to $\Delta \chi^2 \sim 30$ at 68\% C.L., vs $\Delta \chi^2=1$ used in various other PDF fits
adopting the Hessian approach, although not in all~\footnote{MSHT, for instance, uses a dynamical tolerance procedure~\cite{Martin:2009iq}, while CT18 uses a combination of global and dynamic tolerance.}), and this prevents any strong conclusion. The CT18 collaboration has investigated the impact of first SeaQuest data of Ref.~\cite{SeaQuest:2021zxb} on their NNLO PDFs in Ref.~\cite{Park:2021kgf} and they have compared their predictions even to the BNL STAR data on $W$-boson production~\cite{STAR:2020vuq}. Additionally, the CT18A variant of the fit, together with further variants incorporating lattice QCD data on the strangeness asymmetry distribution $s(x) - \bar{s}(x)$, have also been compared to first SeaQuest data of Ref.~\cite{SeaQuest:2021zxb} in Ref.~\cite{Hou:2022onq}.
An advanced study aiming at separating the so-called connected and disconnected sea components, reflecting the topology of the quark lines in the four-point current-current correlator in the nucleon, under the CT18 parameterization, has led to the CT18CS fit~\cite{Hou:2022ajg}, using as a basis the original CT18 data sets. The CT18CS PDFs have also been compared with the distributions extracted from the SeaQuest data of Ref.~\cite{SeaQuest:2021zxb}, and older E866 data of Ref.~\cite{Towell:2001nh}. 

On the other hand, a comparison of the SeaQuest data with 
predictions obtained with the MSHT20~\cite{Bailey:2020ooq} and
CJ15~\cite{Accardi:2016qay} fits, both shown in Fig.~\ref{fig:seaquest},
reveals that the $\bar{d}/\bar{u}$ ratio according to the latter has a trend
compatible with the data only in part of the $(x_1, x_2)$ range.
The CJ collaboration has also investigated the impact of the first SeaQuest data of Ref.~\cite{SeaQuest:2021zxb} plus the aforementioned STAR data
on the CJ15 PDFs in Ref.~\cite{Park:2021kgf}~\footnote{Almost simultaneously, Ref.~\cite{Cocuzza:2021cbi} has presented a global QCD analysis using these same data in the JAM Bayesian Monte Carlo framework.} 
and very recently proposed the new global PDF fit CJ22 in a follow-up paper~\cite{Accardi:2023gyr}, incorporating the SeaQuest data plus the aforementioned STAR data, including higher-twist effects and nucleon off-shell corrections. 
It would be interesting to study as well the modification of the MSHT20
fit, after inclusion of the SeaQuest data~\footnote{Preliminary results in this direction have been shown in talks by the MSHT20 collaboration
at the DIS 2022 and~2023
International Workshops on Deep-Inelastic Scattering and Related Subjects}. 

Finally, we remark that the behaviour of the $(\bar{d}/\bar{u})(x)$ ratio
predicted by the ATLAS 2016 fit~\cite{ATLAS:2016nqi} turns out to be incompatible with the SeaQuest
data, systematically underestimating the latter, pointing to issues in the parameterization
of these PDFs and/or shortcomings during the fit. 
In particular, the comparison of the $x_F$ distribution with the SeaQuest data in Fig.~\ref{fig:seaquest} confirms the point raised 
already in
Ref.~\cite{Alekhin:2017olj} that the assumptions concerning $d$-quark
suppression with respect to the $u$-quarks 
in the ATLAS PDF parameterization adopted in that fit, now outdated, 
are problematic~\footnote{
Also observe that in the CJ15 PDFs the $d$-quark content of the proton is parameterized in terms of the $u$-quark one, introducing a correlation that can affect results for the $d/u$ ratios at large $x$'s, as discussed in Ref.~\cite{Alekhin:2022tip, Alekhin:2022uwc}.}. 
The considerations in Ref.~\cite{Alekhin:2017olj} were based on the
observation that these PDFs already exhibited disagreement with the E866 data,
which were already accessible at that particular time.
One should in any case not be surprised that this old PDF fit is not
in agreement with SeaQuest data, considering that, by definition, it did not include typical non-ATLAS datasets constraining high-$x$ PDFs. In turns, this lack of data required to make more constraining assumptions on the PDF form. Newer ATLAS PDF fits have added more ATLAS data, partially extending $x$ coverage and allowing for more fle\-xi\-ble parameterizations.  
However, we have verified that even the central PDF from a more recent ATLAS fit, ATLASepWZVjet20-EIG~\cite{ATLAS:2021qnl} (not shown in our plot), including the $W, Z/\gamma^*$+jet data that are sensitive to partons at larger $x$'s than the inclusive $W, Z/\gamma^*$ data, turns out to
be also incompatible with the SeaQuest data, overestimating the data up to
several ten percent in the smallest $x_F$ bin (corresponding to the largest $x$). On the other hand, the central PDF from a most recent ATLAS PDF fit, ATLASpdf21~\cite{ATLAS:2021vod} (not shown in our plot), a fit that has included further data and also considered the role of scale uncertainties, 
largely overestimates the SeaQuest data in the first $x_F$ bin, but is compatible with the latter in the other bins, i.e. for $0.2 < x_F < 0.8$. The lack of agreement in the smallest $x_F$ bin can be probably attributed to the fact that ATLAS does not have data constraining $\bar{u}(x)$ and $\bar{d}(x)$ for $x >$ 0.3. On the other hand, the agreement visible at larger $x_F$, corresponding to $x <$~0.3, remarks the compatibility between SeaQuest and DY ATLAS and Tevatron data.
In Ref.~\cite{ATLAS:2021vod} the ATLAS collaboration has provided their own comparison of ATLASpdf21 $\bar{d}/\bar{u}(x)$ ratio with that extracted by the NuSea and SeaQuest collaborations in Ref.~\cite{Towell:2001nh,SeaQuest:2021zxb}. Considering that tha smallest $x_F$ correspond to the largest $x$ values, our results and conclusions on compatibility between fixed-target and collider DY datasets are compatible with their ones.

\section{Impact of nuclear corrections}
\label{sec:nuclear}

SeaQuest data discussed and used in previous sections have been collected for
a deuteron target and for this reason the analysis should address the
corresponding nuclear corrections. 
Here we discuss the effect of nuclear corrections on the DY cross section following Ref.~\cite{Kulagin:2014vsa}.
This model addresses a number of mechanisms for nuclear corrections including
the effect of nuclear momentum distribution (Fermi motion), nuclear binding,
the off-shell modification of bound nucleon PDFs, as well as meson-exchange currents and nuclear shadowing corrections. 
For the kinematics of SeaQuest data the relevant corrections originate from nuclear momentum distribution, binding and off-shell effects on the PDFs.
The deuteron PDFs $q_{i/d}$ of type $i=u,d,\ldots$ can be written as follows~\cite{Kulagin:2014vsa}
(see also Appendix B of Ref.~\cite{Alekhin:2022tip})
\begin{equation}\label{eq:dconv}
xq_{i/d}(x,Q^2)=\int d^3\bm k \left|\Psi_d(\bm k)\right|^2 (1+k_z/M)
x'\left(q_{i/p}(x',Q^2,k^2)+q_{i/n}(x',Q^2,k^2)\right),
\end{equation}
where $q_{i/p(n)}$ is the corresponding proton (neutron) PDFs,
the integration is performed over the nucleon momentum $\bm k$,
$\Psi_d(\bm k)$ is the deuteron wave function in the momentum space,
which is normalized as $\int d^3\bm k \left|\Psi_d(\bm k)\right|^2=1$,
and $M$ is the nucleon mass.
We consider the deuteron in the rest frame and the $z$ axis is chosen to be
antiparallel to the momentum transfer.
The four-momentum of the bound nucleon is $k=(M_d-\sqrt{M^2+\bm k^2}, \bm k)$,
where $M_d$ is the deuteron mass and
$k^2=k_0^2-\bm k^2$ is the invariant mass squared (virtuality), while 
$x'=xM/(k_0+k_z)$ is the Bjorken variable of the off-shell nucleon.

It is convenient to discuss the virtuality dependence of the nucleon PDFs in terms of the
dimensionless variable $v=(k^2-M^2)/M^2$. Since nuclei are weakly bound systems,
the value of $|v|$ is small on average.
For this reason the off-shell PDFs can be expanded in a power series in $v$ about $v=0$~\cite{Kulagin:1994fz}.
Keeping the terms linear in $v$ we have~\cite{Kulagin:2004ie}
\begin{align}
\label{eq:off1}
q(x,Q^2,k^2) &= q(x,Q^2)[1+\delta f(x,Q^2)\, v],
\\
\label{eq:deltaf}
\delta f (x,Q^2) &= \partial\ln q(x,Q^2,k^2)/\partial\ln k^2,
\end{align}
where the derivative is taken for $k^2=M^2$.
The function $\delta f (x, Q^2)$ measures the modification of the nucleon PDFs in the off-shell region.
In Eq.~(\ref{eq:off1}), in order to simplify notations, we suppress the subscripts referring to the PDF type $i$.
Also, we implicitly assume an average over the proton ($p$) and neutron ($n$), $q_i=(q_{i/p}+q_{i/n})/2$,
since Eq.~(\ref{eq:dconv}) for the deuteron depends only on this isoscalar PDF combination.
Detailed studies of nuclear DIS, DY lepton-pair and $W/Z$ boson production indicate that the data are consistent with an universal function $\delta f(x)$, independent of the parton type, and without significant scale and nucleon isospin dependencies~\cite{Kulagin:2004ie,Kulagin:2010gd,Kulagin:2014vsa,Ru:2016wfx,Alekhin:2017fpf,Alekhin:2022tip,Alekhin:2022uwc}~\footnote{ 
The proton-neutron asymmetry  $\delta f_p(x) - \delta f_n(x)$ was
constrained in Ref.~\cite{Alekhin:2022uwc} in a global PDF fit using
data on the proton, $^2$H, $^3$H, and $^3$He targets. This asymmetry
is consistent with zero within uncertainties.
}.
In this work we use the results on the function $\delta f (x)$ from the recent
analysis of some of us in Ref.~\cite{Alekhin:2022uwc}.

\begin{figure}[h!]
\begin{center}
\includegraphics[width=0.9\textwidth]{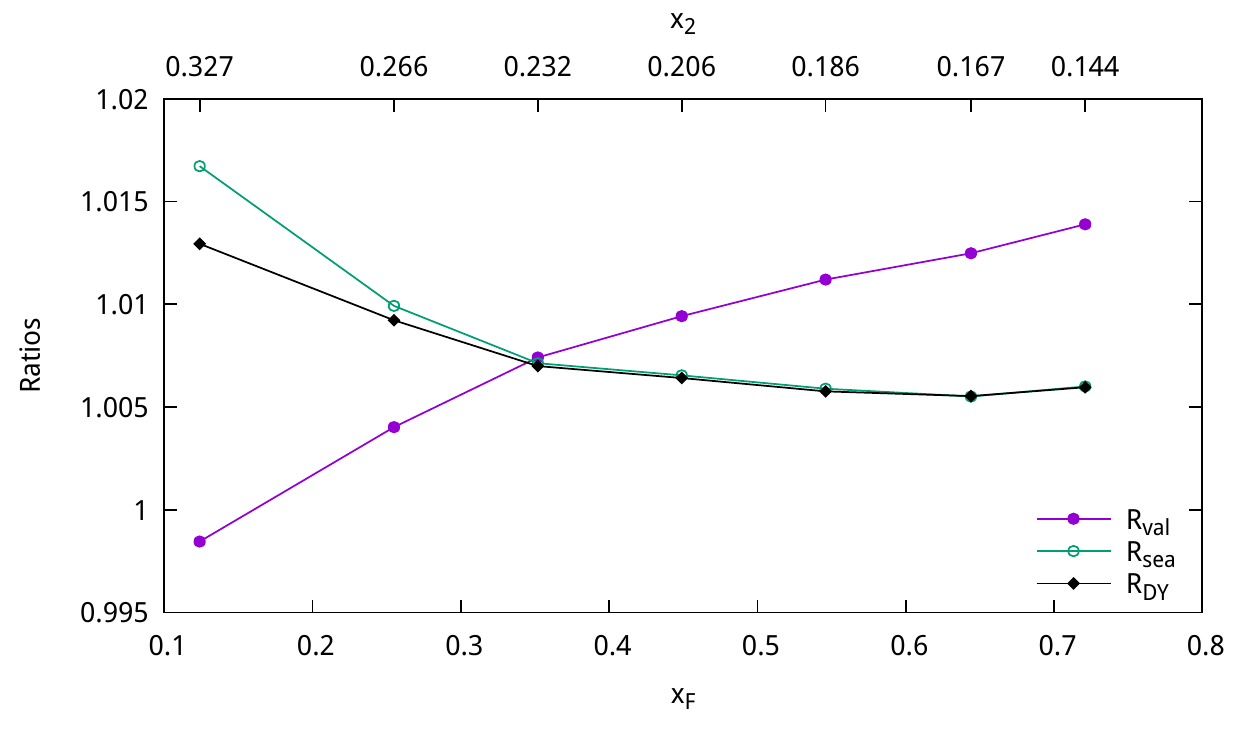}
\caption{\small
\label{fig:nuclearratios} 
Nuclear effects in the deuteron for the valence quark PDFs ($R_{\rm val}$),
antiquark PDFs ($R_{\rm sea}$) and the DY cross sections ($R_{\rm DY}$) (see
text for more detail on the definition of these ratios) vs $x_F$, computed
using Tab.~6 of Ref.~\cite{SeaQuest:2022vwp}. 
The upper horizontal axis indicates the corresponding 
$x$ values of the deuteron target ($x_2$).
}
\end{center}
\end{figure}

In Fig.~\ref{fig:nuclearratios} we illustrate the nuclear effects obtained for
the valence quark PDFs, antiquark PDFs and the DY cross sections for the
kinematics of the SeaQuest experiment. 
In particular, we show the ratios
$R_{\rm val}=u_{{\rm val}/d}/(u_{{\rm val}/p}+u_{{\rm val}/n})$,
$R_{\rm sea}=\overline{u}_{d}/(\overline{u}_p+\overline{u}_n)$ and
$R_{\rm DY}$~=~$\sigma_{pd}/(\sigma_{pp}+\sigma_{pn})$
computed using Eq.~(\ref{eq:dconv}), the NNLO proton PDFs of
Ref.~\cite{Alekhin:2017kpj} and the values of kinematical variables from
Tab.~6 of Ref.~\cite{SeaQuest:2022vwp}. 
Note the different shapes of $R_{\rm val}$ and $R_{\rm sea}$ vs $x_F$.
This is caused by different $x$ dependencies of the valence and antiquark
nucleon PDFs and the smearing effect in the nuclear convolution,
Eq.~(\ref{eq:dconv}). 
The shape and the magnitude of $R_{\rm DY}$ and $R_{\rm sea}$ are similar
corresponding to the fact that the DY cross sections $\sigma_{pd}$, $\sigma_{pp}$ and $\sigma_{pn}$  for SeaQuest kinematics are dominated by the partonic contribution involving a proton beam valence $u$ quark and a target $\bar{u}$, considering PDF $x$ and flavour dependence. However, this dominance is violated for small values of $x_F$  ($x_F < 0.3$), causing the different values of nuclear corrections for DY cross sections and the up-quark sea PDFs in this region.
The  magnitude of the nuclear corrections on the DY $\sigma_{pd}$ is extremely modest, typically 
$\mathcal{O}(0.5 - 1)$\%, and has a 
practically negligible
impact on the present analysis.  
This result is consistent with the claim of Ref.~\cite{SeaQuest:2022vwp} that
nuclear corrections can be neglected, on the basis of the results of
Refs.~\cite{Kamano:2012yx, Ehlers:2014jpa}.

Nuclear corrections should also be
addressed when dealing with data from FNAL-E605 experiment
on proton-copper collisions~\cite{Moreno:1990sf}.
The corresponding corrections on the DY cross sections have been
calculated in Ref.~\cite{Kulagin:2014vsa} (see Fig.~8 and Table~2 
there).
The rate of nuclear corrections depends on both the target $x_2$ and
the mass of the muon pair 
as illustrated in Fig.~8 of Ref.~\cite{Kulagin:2014vsa}.
Note, however, that the E605 experiment only provides data 
for copper target
and did not take data
for the proton target. 
Since copper is almost an isoscalar target with about 8\% of the
neutron excess, the E605 cross section data on copper target alone 
provides a little
sensitivity to measuring the $(\bar{d}-\bar{u})(x)$ asymmetry 
of the sea distributions.
We also verified that removing the E605 data from our fits does not
essentially change the results presented in Figs.~\ref{fig:udm} and
\ref{fig:udmratio}.

\section{Second Mellin moments of quark distributions: comparisons with lattice QCD computations}
\label{sec:lattice}

Important information on PDFs can also be gained from lattice QCD, which gives access to some of their moments.
Recalling that $q(x, Q^2)=q_{\rm val}(x, Q^2)+q_{\rm sea}(x, Q^2)$, $\bar{q}(x, Q^2) = \bar{q}_{\rm sea}(x, Q^2)$, the following definition allows to summarize the 
 moments of the $q^+$~$\equiv$~ $q$~$+$~$\bar{q}$ (total) and $q^-$~$\equiv$~ $q$~$-$~$\bar{q}$ (valence) quark combinations at a scale $Q^2$: 
\begin{equation}
\langle x^{n-1} \rangle_{q^{\pm}} (Q^2)\,\, =  \int_0^1 dx \, x^{n-1} \, q^\pm(x,Q^2)
 \, ,
\end{equation}
where $n=$~1,~2,~3, ... refers to the first, second, third, etc. Mellin moment (e\-qui\-va\-lent to zeroth, first, second, etc. $x$ moment), respectively. 
The first Mellin moments $\langle 1 \rangle_{q^-}$ correspond to the quark number sum rules~\footnote{On the other hand, the moments $\langle 1 \rangle_{q^+}$ are not constrained by symmetries and are divergent.}. 
Lattice QCD computations have allowed to calculate
the second Mellin moments  $\langle x \rangle_{u^+ - d^+}$ (isovector combination) and $\langle x \rangle_{q^+}$ for all individual light quarks,  
together with 
the third Mellin moments $\langle x^2 \rangle_{u^- - \, d^-}$ and $\langle x^2 \rangle_{q^-}$~\cite{Lin:2017snn, Constantinou:2020hdm, Davoudi:2022bnl}. 

In Ref.~\cite{Alekhin:2017kpj} we compared 
the values of $\langle x^2 \rangle_{u^-}$, $\langle x^2 \rangle_{d^-}$, 
$\langle x^2 \rangle_{u^- - \, d^-}$ and $\langle x \rangle_{u^+ - \, d^+}$  
that we computed for various NNLO PDF fits with corresponding values
extrapolated from lattice QCD computations. 
In this work, we update and extend the comparison of Ref.~\cite{Alekhin:2017kpj}. 
On the one hand, in addition to several modern NLO and NNLO PDF fits, we
incorporate the newly presented PDF fits from the previous sections, which
take into account the SeaQuest data. 
As discussed in those sections, these data have minimal impact on the valence quark distributions. 
However, they play a crucial role in constraining the isospin asymmetry $(\bar{d} - \bar{u})(x)$ of the sea-quark distributions. 
On the other hand, we also consider updated evaluations from lattice QCD,
utilizing new computational methods that yield reduced uncertainties compared to previous analyses. 
Additionally, we incorporate the recently released results on the moments of the $u^+$ and $d^+$ distributions, which
were not available at the time of Ref.~\cite{Alekhin:2017kpj}. 

\begin{table}[h!]
\begin{center}
\footnotesize
\begin{tabular}{l|l|l|l}
\hline
\textbf{PDF fit} &
${ \langle x \rangle_{u^+}}$ &
$ \langle x \rangle_{d^+}$ &
$ \langle x \rangle_{u^+ - d^+}$ 
\\
\hline
ABMP16 + SeaQuest NLO 
    &\, 0.3523 $\pm$ 0.0010
    &\, 0.1813 $\pm$ 0.0023
    &\, 0.1711 $\pm$ 0.0029\\
ABMP16 + SeaQuest NNLO 
    &\, 0.3535 $\pm$ 0.0026
    &\, 0.1858 $\pm$ 0.0028
    &\, 0.1677 $\pm$ 0.0036    \\
ABMP16  NLO 
    &\, 0.3522 $\pm$ 0.0026  
    &\, 0.1814 $\pm$ 0.0027 
    &\, 0.1708 $\pm$ 0.0036 \\ 
ABMP16 NNLO 
    &\, 0.3532 $\pm$ 0.0027  
    &\, 0.1858 $\pm$ 0.0029 
    &\, 0.1673 $\pm$ 0.0037 \\ 
NNPDF4.0 NNLO 
    &\, 0.3468 $\pm$ 0.0026  
    &\, 0.1934 $\pm$ 0.0032 
    &\, 0.1533 $\pm$ 0.0041 \\
CT18 NNLO 
    &\, 0.3498 $^{+\, 0.0078}_{-\, 0.0085}$ 
    &\, 0.1934 $^{+\, 0.0083}_{-\, 0.0103}$ 
    &\, 0.1564 $^{+\, 0.0123}_{-\, 0.0120}$ \\ 
MSHT20 NNLO 
    &\, 0.3471 $^{+\, 0.0048}_{-\, 0.0048}$  
    &\, 0.1923 $^{+\, 0.0046}_{-\, 0.0060}$ 
    &\, 0.1548 $^{+\, 0.0062}_{-\, 0.0056}$ \\
CJ15 NLO 
    &\, 0.3480 $^{+\, 0.0009}_{-\, 0.0012}$  
    &\, 0.1962 $^{+\, 0.0015}_{-\, 0.0014}$ 
    &\, 0.1518 $^{+\, 0.0019}_{-\, 0.0024}$ \\
epWZ16 NNLO 
    &\, 0.3628 $^{+\, 0.0027}_{-\, 0.0028}$ 
    &\, 0.1741 $^{+\, 0.0047}_{-\, 0.0039}$ 
    &\, 0.1887 $^{+\, 0.0041}_{-\, 0.0050}$
 \\
\hline
\hline
\textbf{lattice computation}  &  &  &  \\
$\chi$QCD18~\cite{Yang:2018nqn} ($n_f = 2+1$) 
    &\, 0.307 $\pm$ 0.030 $\pm$ 0.018 
    &\, 0.160 $\pm$ 0.027 $\pm$ 0.040 
    &\,  0.151 $\pm$ 0.028 $\pm$ 0.029 \\
RQCD18~\cite{Bali:2018zgl} ($n_f = 2$) 
    &$\quad$ - 
    &$\quad$ - 
    &\, 0.195 $\pm$ 0.007 $\pm$ 0.015 \\
ETMC20~\cite{Alexandrou:2020sml} ($n_f = 2 + 1 + 1$) 
    &\, 0.359 $\pm$ 0.030 
    &\, 0.188 $\pm$ 0.019 
    &\, 0.171 $\pm$  0.018 \\
PNDME20~\cite{Mondal:2020cmt} ($n_f = 2 + 1 + 1$) 
    &$\quad$ - 
    &$\quad$ - 
    &\, 0.173 $\pm$  0.014 $\pm$ 0.007 \\
NME20~\cite{Mondal:2020ela} ($n_f = 2 + 1$) 
    &$\quad$ - 
    &$\quad$ - 
    &\, 0.155 $\pm$ 0.017  $\pm$ 0.020 \\
Mainz21~\cite{Ottnad:2021tlx} ($n_f = 2 + 1$) 
    &$\quad$ - 
    &$\quad$ - 
    &\, 0.139 $\pm$  0.018 ({$\mathrm{stat}$}) \\
\hline
\end{tabular}
\caption{\small
\label{tab:secondmom}
Comparison of second Mellin moments for various combinations of light-quark distributions from different NLO and NNLO PDF fits, including those proposed in this work, with uncertainties due to PDF variations, 
to corresponding values extrapolated from $n_f$-flavour lattice QCD computations ($Q$~=~2~GeV). In the case of the CT18 fit, the uncertainties refer to the 90\% C.L. interval, instead of the 68\%  one used by other Hessian PDF fits. 
}
\end{center}
\end{table}

\begin{figure}[h!]
\begin{center}
\includegraphics[width=0.3154\textwidth]{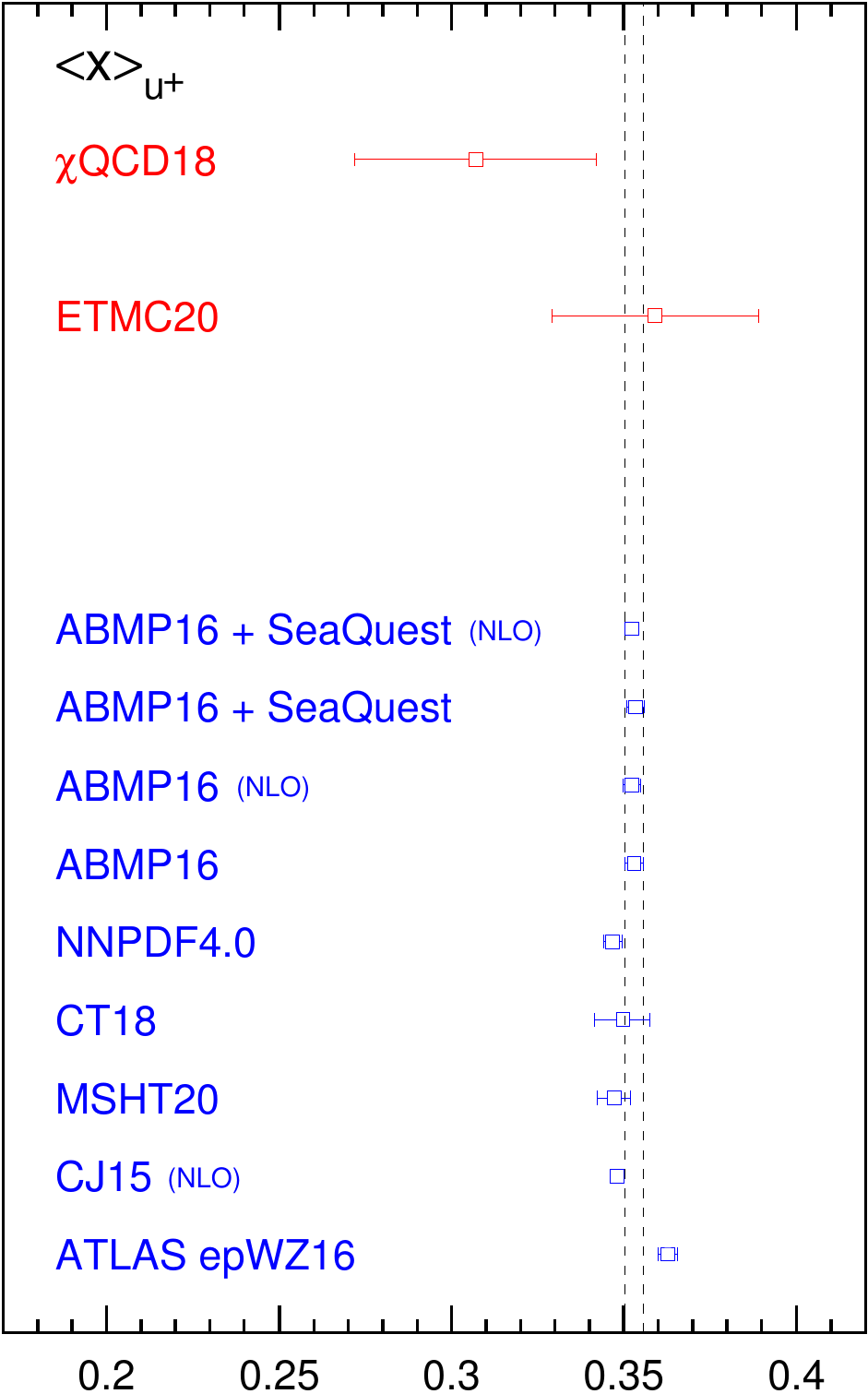}
\hspace{0.10cm}
\includegraphics[width=0.33\textwidth]{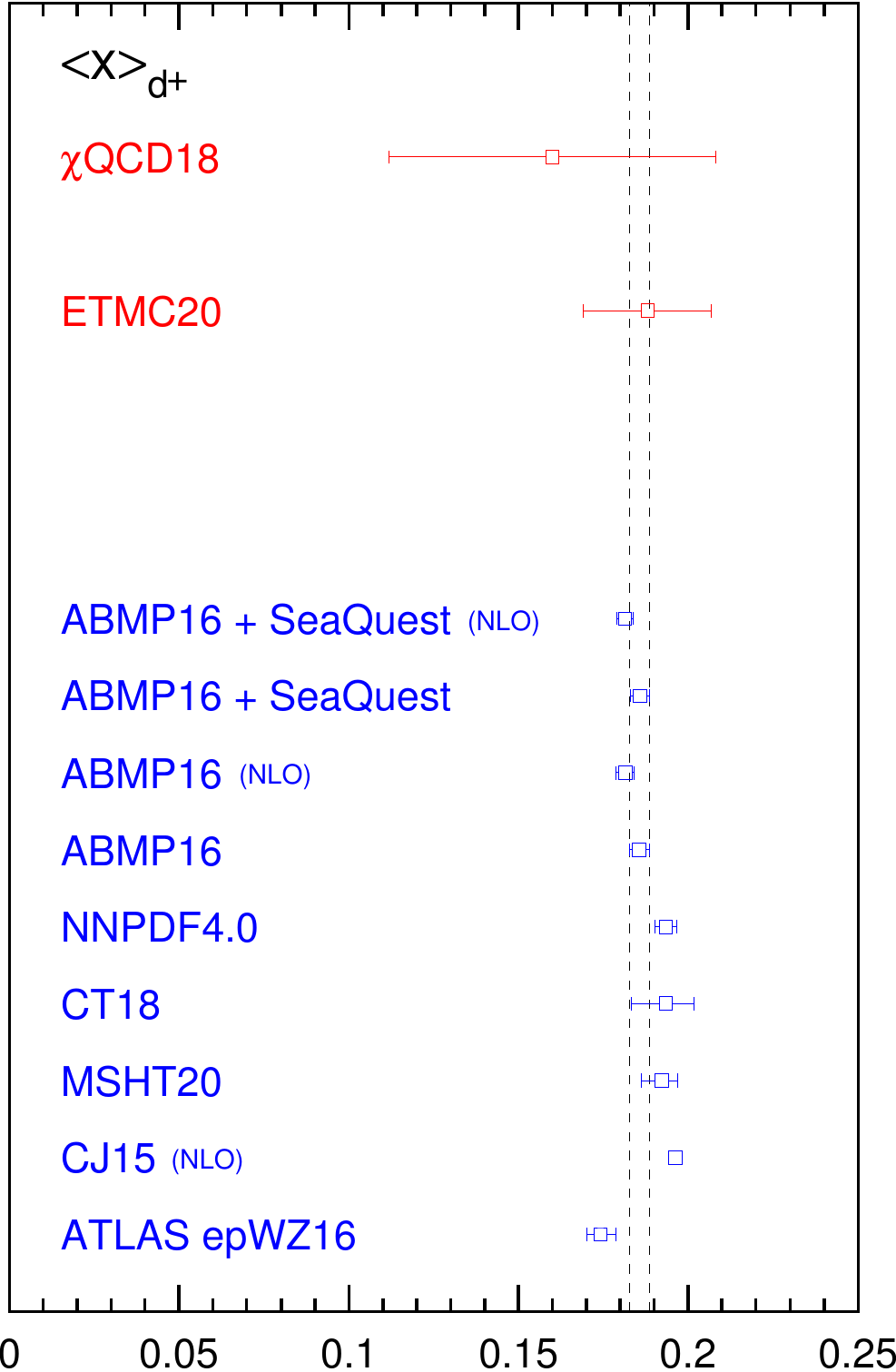}
\includegraphics[width=0.33\textwidth]{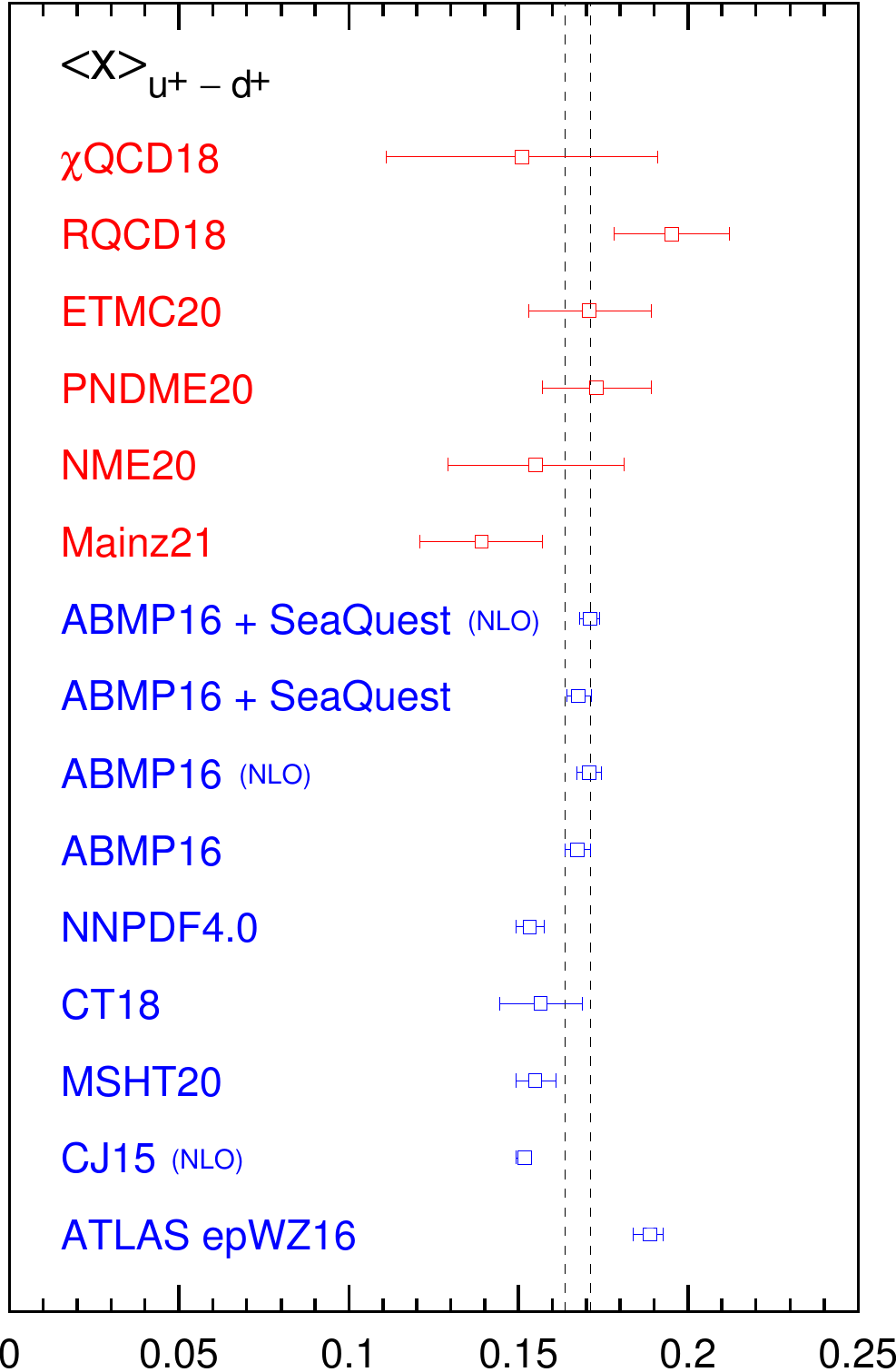}
\caption{\small
\label{fig:secondmom} 
Second Mellin moments of $u^+(x)$, $d^+(x)$ and the isovector combination $(u^+ - d^+)(x)$ and their uncertainties computed for a range of PDF fits and from lattice QCD. The corresponding numerical va\-lues are tabulated in the columns of Tab.~\ref{tab:secondmom} and reported in the panels of this plot for a more immediate visualization. The vertical band in each panel brackets the values from the ABMP16 NNLO fit. 
}
\end{center}
\end{figure}

In Tab.~\ref{tab:secondmom} we compare our calculations of second Mellin moments
using as a basis the NLO and NNLO quark distributions considered in the previous section, 
to the
most recent results from lattice QCD~\cite{Yang:2018nqn, Bali:2018zgl,
  Alexandrou:2020sml, Mondal:2020cmt, Mondal:2020ela, Ottnad:2021tlx}. 
In par\-ti\-cu\-lar, the $\chi$QCD and ETMC collaborations have recently released
data on the second moments of $u^+$, $d^+$ and $u^+ - d^+$ combinations
dependent on both valence and sea quarks, in Ref.~\cite{Yang:2018nqn}
and~\cite{Alexandrou:2020sml}, respectively, while the RQCD, NME, PNDME and
Mainz collaborations have determined the second moments of the $u^+ - d^+$ combination in Ref.~\cite{Bali:2018zgl, Mondal:2020cmt, Ottnad:2021tlx},
respectively~\footnote{In the case where lattice collaborations have released values of second moments in more than one work, we only cite the most updated ones.}. 
The outcomes from Tab.~\ref{tab:secondmom} are also
represented graphically in Fig.\ref{fig:secondmom}.

We observe that the present status of PDF fits is advanced to the point that
the quoted uncertainties on the second Mellin moments, which represent the
experimental data uncertainties pro\-pa\-ga\-ted through the fit, are so small 
that the results from different fits are not always compatible among each other
within their uncertainties. 
This is partly related to the theory assumptions made in those fits, but
also 
due to the data sets considered, i.e., inclusion of DY data from colliders, see e.g., Ref.~\cite{Accardi:2016ndt}.
Across different orders of perturbation theory, the second Mellin moment values and uncertainties from NLO and NNLO PDF fits have almost comparable values, 
indicating very good perturbative stability.

The corresponding second Mellin moments from QCD lattice computations turn out to be significantly more uncertain and not yet able to discriminate between the various PDF fits. 
Taking into account the range of lattice results and the inherent uncertainty
associated with each of them, they are presently exhi\-bi\-ting a high level of
compatibility with nearly all
the PDF fits.
The lattice moments $\langle x \rangle_{u^+}$ by the $\chi$QCD collaboration 
 and $\langle x\rangle_{u^+ -\,d^+}$ by the RQCD collaborations,  
 both computed in 2018, exhibit a slight tension, deviating from their 1$\sigma$
range, when compared to the majority of PDF fits. Nonetheless, these results
carry substantial uncertainties and align with the findings of PDF fits within a
2$\sigma$ range.
Most of the recent lattice results, in particular those obtained by the ETMC
collaboration in 2020, turn out to agree very well with almost all PDFs fits. 
The 2021 result on $\langle x \rangle_{u^+ - \, d^+}$ of the Mainz collaboration agrees with moments of some of the global PDF fits, but
is slightly smaller, although compatible within $2\sigma$, with the second moments from the ABMP16~(+~SeaQuest) NLO and NNLO PDFs.  

We also observe that the addition of SeaQuest data to the ABMP16 NNLO PDF fit has a tiny effect on the values of the considered moments, slightly decreasing the associated uncertainties, while the central values remain approximately the same. 
The improvement of the uncertainties turns out to be more pronounced in the case of the ABMP16 NLO PDF fit. 
Overall, the results from NLO and NNLO ABMP16~(+~SeaQuest) PDFs are consistent
among each other and, as mentioned, the order of perturbation theory does not
have a significant impact on the second Mellin moments, being rather inclusive quantities.

In light of the comparisons discussed here, it will be interesting to observe
precise lattice calculations of higher Mellin moments (beyond the second/third ones), 
maybe exploiting concepts and techniques of Ref.~\cite{Detmold:2005gg, Davoudi:2012ya}, so as to enable similar comparisons for the fourth, etc. moments.  Another valuable improvement would be the ability to distinguish between
valence and sea quark PDFs in lattice results.

\section{Conclusions}
\label{sec:conclu}

We have studied a variant of the ABMP16 NLO and NNLO fits, including the
SeaQuest non-resonant data on $\sigma_{pd}\,/ (2 \sigma_{pp})$ as a function of $x_F$.
We find that these data reduce uncertainties on the $(\bar{d} - \bar{u})(x)$
difference as well as on the $(\bar{d}/\bar{u})(x)$ ratio at large $x$, 
while leaving essentially unchanged the values of the other quantities, which are simultaneously
constrained in these fits ($\alpha_s(M_Z)$ and heavy-quark masses).
The $\chi^2/\mathrm{NDP}$ for the fits including SeaQuest data are within
statistical uncertainty of those previously obtained without these data. 
The simultaneous description of all DY data turns out to be slightly more
consistent at NNLO than at NLO, as expected for the improved precision of the
theoretical predictions.
In particular, we observe the compatibility of SeaQuest data constraints on the $\bar{d}-\bar{u}$ asymmetry, with the corresponding constraints from collider DY data at both the Tevatron and the LHC. 
This confirms the presence of an asymmetric sea, ruling out PDF fits based on the  assumption of (or leading to) a symmetric sea.

Our present results support using the SeaQuest data together with collider
DY data in future updated 
PDF analyses, that would allow further reducing PDF uncertainties, as well as a cross-check of the compatibility with the data already included there. The inclusion of SeaQuest data is facilitated by the fact that nuclear corrections for the deuteron
target, that we have explicitly computed in this work, 
turned out to be $\mathcal{O}(0.5 - 1)$\% in all SeaQuest $x_F$ bins,
thus having a practically negligible effect on the final PDFs. 
The smallness of the observed nuclear effects can be attributed to 
the kinematics of
the SeaQuest experiment itself. The experiment combines partons with
relatively small but still significant $x_2$ values (specifically, $x_{\rm
  target}$) and larger $x_1$ values (specifically, $x_{\rm beam}$), with only
the target experiencing nuclear corrections. The most substantial corrections
occur in the bin with the smallest $x_F$, which corresponds to the largest
$x_2$ values ($\le 0.45$). It is worth noting that larger nuclear corrections
would be anticipated at larger $x_2$ values, which correspond to backward
kinematics that fall outside the scope of SeaQuest's current detector
capabilities.

The second moments of various combinations of light-quark distributions from NLO and NNLO PDF fits
are compatible with current lattice QCD results. Although lattice QCD is not
yet competitive for distinguishing between different PDF fits, advancements in
techniques and increased attention from the lattice community are expected to
improve this limitation in the future.

We strongly encourage the SeaQuest collaboration to continue their efforts in
reducing the uncertainties associated with their measurements, aiming for
values below the current level of approximately 5\%. 
Achieving this would significantly enhance the constraining power of the data
on sea quark distributions. 
It is worth noting that only around one half of the experiment's data has been
utilized for the published studies thus far, suggesting the potential for
further improvements. 
Moreover, it would be highly beneficial if the SeaQuest collaboration released
separate data on $pp$ and $pd$ cross-sections. 
Such separate data sets would enable more precise constraints to be obtained
for the $\bar{u}$ and $\bar{d}$ quark distributions, facilitating a deeper
understanding of their individual characteristics.

\subsection*{Acknowledgements}
We would like to thank R.~Petti for stimulating disussions, 
and S.~Collins and G.~Bali for insightful comments concerning lattice QCD results.
We thank A.~Cooper-Sarkar and F.~Giuli for remarks on the ATLAS PDF fits and the participants of the Les Houches Workshop PhysTeV 2023 for further comments. We are also grateful to the anonimous Referee for the constructive report and for having brought to our attention further studies closely related to the topic of this manuscript. The work of S.A. and S.-O.M. has been supported in part by 
the Deutsche Forschungsgemeinschaft through the Research Unit FOR 2926, {\it Next
  Generation pQCD for Hadron Structure: Preparing for the EIC}, project number 40824754.
The work of M.V.G. and S.-O.M. has been supported in part by
the Bundesministerium f\"ur Bildung und Forschung under contract 05H21GUCCA.

\bibliographystyle{JHEP}
\bibliography{seaquest2}  
\end{document}